\documentclass[12pt,a4paper]{article}
\usepackage{amsmath}
\usepackage{amssymb}
\usepackage{colordvi}
\usepackage{epsfig}
\usepackage{subfigure}
\usepackage{mathrsfs}
\usepackage{cite}

\newcommand{\eq}[1]{Eq.~(\ref{#1})}
\newcommand{\eqs}[1]{Eqs.~(\ref{#1})}

\newcommand{\V}[1]{V_{#1}^{\phantom{\ast}}}
\newcommand{\Vc}[1]{V_{#1}^{\ast}}

\newcommand{\im}[1]{{\text{Im}\left[#1\right]}}
\newcommand{\re}[1]{{\text{Re}\left[#1\right]}}

\newcommand{\DMBs}{\Delta M_{B_s}}
\newcommand{\DMBd}{\Delta M_{B_d}}
\newcommand{\DGBd}{\Delta \Gamma_{d}}
\newcommand{\DGBs}{\Delta \Gamma_{s}}
\newcommand{\DGBsCP}{\Delta \Gamma_{s}^{CP}}
\newcommand{\AJPsi}{A_{J/\psi K_S}}
\newcommand{\BBd}{$B_d^0$--$\bar B_d^0$}
\newcommand{\BBs}{$B_s^0$--$\bar B_s^0$}
\newcommand{\Asld}{A_{sl}^d}
\newcommand{\Asls}{A_{sl}^s}
\newcommand{\Ads}{\mathcal A}
\newcommand{\fBd}{B_{B_d}f_{B_d}^2}
\newcommand{\fBs}{B_{B_s}f_{B_s}^2}
\newcommand{\AJPsiPhi}{A_{J/\Psi\Phi}}
\begin{document}

\begin{flushright}
hep-ph/0608100\\IFIC/06-22\\ FTUV-06-0809
\end{flushright}

\begin{center}
\begin{Large}
{\bf CP violation and limits on New Physics including recent $\boldsymbol{B_s}$ measurements}\\
\end{Large}
\vspace{0.5cm}
Francisco J. Botella$^{~a}$, Gustavo C. Branco$^{~b,c}$, Miguel Nebot$^{~b}$\\ \vspace{0.3cm}
{\small \emph{
$^a$ Departament de F\'{\i}sica Te\`orica and IFIC,\\ Universitat de Val\`encia-CSIC,\\ E-46100, Burjassot, Spain\\
$^b$ Centro de F\'{\i}sica Te\'orica de Part\'{\i}culas (CFTP),\\ Instituto Superior T\'ecnico,\\ P-1049-001, Lisboa, Portugal\\
$^c$ Physik Department, Technische Universit\"{a}t M\"{u}nchen,\\ D-85748, Garching, Germany
}}
\end{center}

\begin{abstract}
We analyse present constraints on the SM parameter space and derive, in a model independent way, various bounds on New Physics contributions to \BBd~ and \BBs~ mixings. Our analyses include information on a large set of asymmetries, leading to the measurement of the CKM phases $\gamma$ and $\bar\beta$, as well as recent data from D0 and CDF related to the \BBs~ system such as the measurement of $\DMBs$, $\Ads$ and $\DGBsCP$. We examine in detail several observables such as the asymmetries $\Asld$, $\Ads$, the width differences $\DGBd$ and $\DGBsCP$ and discuss the r\^ole they play in establishing the limits on New Physics. The present data clearly favour the SM, with the New Physics favoured region placed around the SM solution. A New Physics solution significantly different from the SM is still allowed, albeit quite disfavoured (2.6\% probability). We analyse the presently available indirect knowledge on the phase $\bar\chi$ entering in \BBs~ mixing and study the impact of a future measurement of $\bar\chi$ to be achieved at LHC, through the measurement of the time-dependent CP asymmetry in $B_s\to J/\Psi\ \Phi$ decays.
\end{abstract}

\section{Introduction}
In the past few years, there has been a remarkable progress in flavour physics, both in theory and experiment, with an impressive amount of experimental data which can provide precision tests \cite{Botella:2002fr} on the flavour sector of the Standard Model (SM).

  Perhaps the most fundamental task of experiments on CP violation, was to provide an irrefutable proof that the Cabibbo-Kobayashi-Maskawa \cite{Cabibbo:1963yz,Kobayashi:1973fv} matrix is non-trivially complex, thus implying that charged weak interactions do violate CP. This task has been achieved with the recent measurements \cite{Aubert:2004zt,Sciolla:2005kz,Aubert:2004kv,Aubert:2006ia,Aubert:2003wr,Aubert:2004zr,Aubert:2005nj,Aubert:2006af,Aubert:2004cp,Aubert:2005yf,Aubert:2006tw,Abe:2005bt,Abe:2004gu,Poluektov:2006ia,Zhang:2003up,Somov:2006sg,Itoh:2005ks,Abe:2004mu} of the angles $\gamma$ and $\bar\alpha$ \cite{Gronau:1990ra,Gronau:1991dp,Dunietz:1991yd,Aleksan:1991nh,Atwood:1994zm,Atwood:1996ci,Giri:2003ty,Gronau:1990ka,Grossman:1997jr,Charles:1998qx,Gronau:2001ff,Snyder:1993mx,Botella:2005ks} which provide clear evidence \cite{Botella:2005fc} for a complex CKM matrix even if one allows for the presence of essentially arbitrary New Physics (NP) contributions at loop level. This is an important result, with profound impact on the question of the origin of CP violation. Let us consider, for example, theories where CP is a good symmetry of the Lagrangian, only broken by the vacuum \cite{Lee:1973iz}. The experimental evidence that the CKM matrix is complex even if one allows for the presence of NP, implies that among theories with spontaneous CP breaking, only those where the vacuum phases also generate a complex CKM matrix while at the same time suppressing Flavour Changing Neutral Currents, are viable \cite{Branco:2006av}. In particular, certain classes  of SUSY extensions of the SM with spontaneous CP breaking \cite{Branco:2000dq,Hugonie:2003yu}, as well some multi-Higgs theories with natural flavour conservation are no longer valid since they lead to a real CKM matrix \cite{Branco:1979pv,Branco:1980sz,Branco:1985pf}. Fortunately, it has been recently shown \cite{Branco:2006pj} that it is possible to have a SUSY extension of the SM with spontaneous CP breaking and a complex CKM matrix. This is achieved through the introduction of two singlet chiral superfields and a vector-like quark chiral superfield, which mixes with standard quarks.  
 
 Another major task for present and future experiments on flavour and CP violation is to either discover or put bounds on NP contributions to flavour mixing and CP violation. At this stage, it should be emphasized that it is clear that there are new sources of CP violation beyond those present in the SM. On the one hand, CP violation present in the SM is not sufficient to generate the observed baryon asymmetry of the universe (BAU), and on the other hand new sources of CP violation are present in essentially all extensions of the SM, including the supersymmetric ones. The important open question is whether these new sources of CP violation will be visible at low energy experiments, or not.
       
In this paper, we analyse present constraints on the SM parameter space and derive bounds on the size of NP contributions, taking into account the recent results on $\DMBs$ as well as on semileptonic asymmetries and on width differences obtained both at B factories and at the Tevatron. In the study of the constraints on NP we assume that tree level decays are dominated by the SM amplitudes but allow for the possibility of significant NP contributions to \BBd~ and \BBs~ mixings and in general to all other SM processes which are only induced at the loop level. We study in detail the r\^ole played by each individual measurement in conforming the allowed regions of NP parameter space.

 The paper is organised as follows. Section \ref{sec:BBd} presents the starting point for the different analyses, i.e. the use of tree level extracted CKM moduli and phases together with arbitrary NP contributions to B mesons mixings within a $3\times 3$ unitary CKM matrix framework. It is then extended to understand the r\^ole that $\Gamma_{12}^d/M_{12}^d$ will play. No information on \BBs~ is used until section \ref{sec:BBs}. In this section we first include recent measurements of $\DMBs$ and then study $\Gamma_{12}^s/M_{12}^s$ and different relevant observables. Section \ref{sec:Complete} presents the results of a complete analysis including all the observables considered before. Finally, we present our conclusions in section \ref{sec:Conclusions}.

\section{NP analyses without the inclusion of $\boldsymbol{B_s^0}$--$\boldsymbol{\bar B_s^0}$ measurements \label{sec:BBd}}

\subsection{Previous situation}
Previous analyses \cite{Ligeti:2004ak,Botella:2005fc,Bona:2005eu,Silvestrini:2005zb,Charles:2004jd} addressing the CP-violating nature of the CKM matrix and New Physics contributions to flavour processes provided several interesting lessons:
\begin{itemize}
\item Using tree level measurements of moduli, namely $|\V{ud}|$, $|\V{us}|$, $|\V{ub}|$, $|\V{cd}|$, $|\V{cs}|$ and $|\V{ub}|$, to reconstruct genuinely CP-violating quantities like the invariant $\text{Im}\ Q\equiv\text{Im}(\V{us}\V{cb}\Vc{ub}\Vc{cs})$, even if \emph{a priori} feasible, was shown to be irrelevant as it would require totally unrealistic precision in the determination of $|\V{us}|$, $|\V{cd}|$. Including $|\V{td}|$ -- obtained form $\DMBd$ -- in this type of analysis is trivial, in case no NP is considered to contribute to \BBd~ mixing, one can indeed derive $\text{Im}\ Q$ from $|\V{us}|$, $|\V{ub}|$, $|\V{cb}|$, $|\V{td}|$. However, this result has the drawback that the presence of NP in the mixing prevents the use of $|\V{td}|$. This would equally apply to the use of $|\V{ts}|$ and $\DMBs$. The complex nature of the CKM matrix was subsequently established, without regard to potential NP in \BBd, through the additional use of $\AJPsi$, $\gamma$ and $\bar\alpha$ measurements.
\item Considering the presence of NP in the \BBd~ mixing and the use of measurements of $\gamma$ and $\AJPsi$, one is lead to find not only a SM-like solution but three additional solutions including significant NP contributions owing to the discrete ambiguities inherent to the determination of $\gamma$ and $\AJPsi$: the measurement of $\gamma$ has a $\pi$ ambiguity, for $\AJPsi$ there is also a twofold ambiguity since $\AJPsi=\sin (2\bar\beta)>0$ gives $2\bar\beta\in [0;\frac{\pi}{2}]$ or $2\bar\beta\in [\frac{\pi}{2};\pi]$. These four solutions cannot be distinguished by a set of observables like $\{$$|\V{ud}|$, $|\V{us}|$, $|\V{ub}|$, $|\V{cd}|$, $|\V{cs}|$, $|\V{cb}|$, $\gamma$, $\AJPsi$, $\DMBd$$\}$.
\item The inclusion of additional observables may be helpful in disfavouring some of the previous solutions. Consider for example $\bar\alpha=\pi-\bar\beta-\gamma$, despite being measured with a $\pi$ ambiguity, it can distinguish the solutions with $2\bar\beta\in [0;\frac{\pi}{2}]$ and those with $2\bar\beta\in [\frac{\pi}{2};\pi]$, i.e. it is sensitive to the sign of $\cos(2\bar\beta)$; however there is no distinction among $\gamma$ and $\gamma+\pi$. The same would apply to the measured phase $2\bar\beta+\gamma$.
\end{itemize}
To summarize, those analyses showed: (i) the CKM matrix is complex beyond any reasonable doubt even if one allows for the presence of NP and (ii) there is room for New Physics in \BBd, either close to SM values (in which case NP does relax \emph{tensions} among different observables not completely consistent in the SM) or with values neatly different from those ones.

\begin{figure}[h]
\begin{center}
\subfigure[Apex of the unitarity triangle $db$, $\im{-\frac{V_{ud}V_{ub}^\ast}{V_{cd}V_{cb}^\ast}}$ vs. $\re{-\frac{V_{ud}V_{ub}^\ast}{V_{cd}V_{cb}^\ast}}$\label{figure01a}]{\epsfig{file=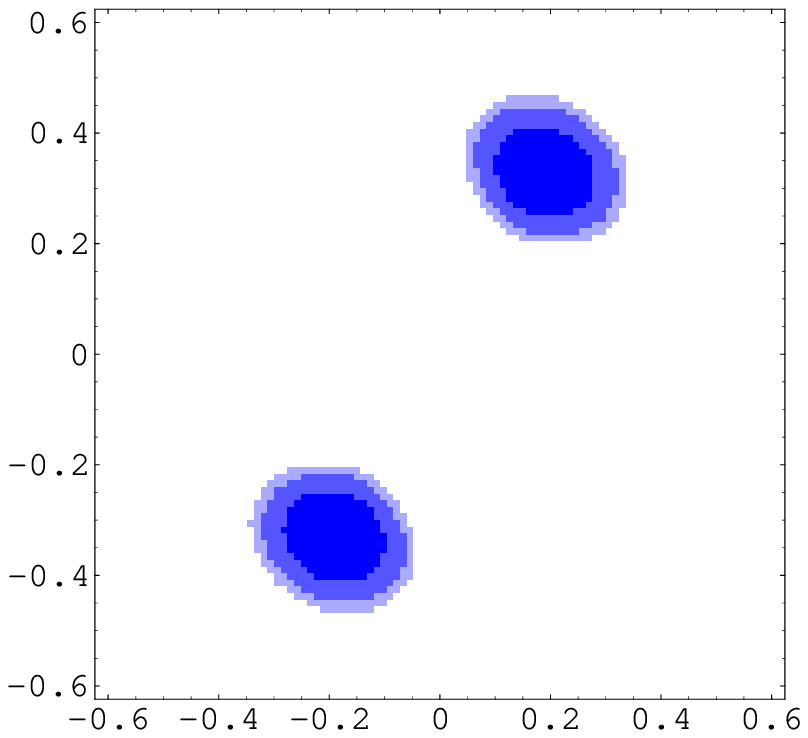,width=0.45\textwidth}}\quad \subfigure[$r_d^2$ vs. $2\phi_d$\label{figure01b}]{\epsfig{file=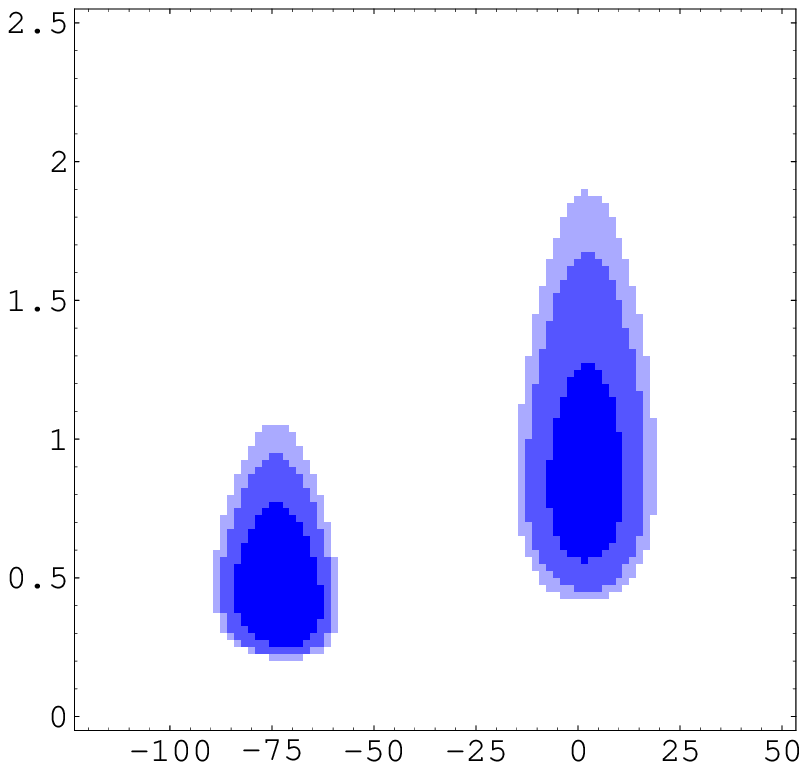,width=0.45\textwidth}}
\caption{Probability regions: $68\%$ dark, $90\%$ medium, $95\%$ light.}\label{figure01}
\end{center}
\end{figure}

This is the starting point of our analyses, illustrated in figure \ref{figure01}. The basic set of constraints used here -- and in several other places along this work -- is given by:
\begin{eqnarray}
|\V{ud}|\quad\quad |\V{us}|\quad\quad |\V{ub}|& |\V{cd}|\quad\quad |\V{cs}|\quad\quad |\V{cb}|\notag\\
\AJPsi\quad\quad \gamma \quad\quad \bar\alpha\quad & 2\bar\beta+\gamma\quad\quad \cos 2\bar\beta\quad\quad \DMBd ~.\label{basicset}
\end{eqnarray}
Numerical values are shown in table \ref{table:inputs} in the appendix.

Among the NP solutions, the use of $\bar\alpha$, $\cos 2\bar\beta$ and $2\bar\beta+\gamma$ strongly disfavours the ones with $\arg(\AJPsi)\sim 138^\circ$ (see table \ref{table01}). In the following, in terms of $\gamma$ and $2\phi_d$, we will mainly focus on the SM-like solution, with $\gamma\sim 65^\circ$ and $\arg(\AJPsi)\sim 42^\circ$, and on the NP solution with $\gamma\sim -115^\circ$ and $\arg(\AJPsi)\sim 42^\circ$.

\begin{table}[h]
\begin{center}
\begin{tabular}{|c|c|c|c|}
\hline
$\gamma$ & $\arg(\AJPsi)$ & $2\phi_d$ & Probability\\ \hline\hline
$65^\circ$ & $42^\circ$ & $3^\circ$ & $\sim 50\%$ \\ \hline
$-115^\circ$ & $42^\circ$ & $-76^\circ$ & $\sim 50\%$ \\ \hline
$65^\circ$ & $138^\circ$ & $-92^\circ$ & $< 0.2\%$ \\ \hline
$-115^\circ$ & $138^\circ$ & $-171^\circ$ & $< 0.2\%$ \\ \hline
\end{tabular}
\caption{Approximate central values for $\gamma$, $2\bar\beta=\arg(\AJPsi)$ and $2\phi_d$ corresponding to the four solutions. The last column shows the corresponding probability.}\label{table01}
\end{center}
\end{table}

\subsection{The r\^{o}le of $\boldsymbol{\Gamma_{12}^d/ M_{12}^d}$: $\boldsymbol{\Asld}$ and $\boldsymbol{\DGBd}$}

In this subsection we will analyse how the inclusion of two additional observables, the semileptonic asymmetry in $B_d$ decays, $\Asld$, and the difference in the widths of the eigenstates of the effective hamiltonian controlling the \BBd~ oscillations, $\Delta\Gamma_d$, can change the picture. Both observables are controlled by the same quantity, $\Gamma_{12}^d/ M_{12}^d$: $\Asld$ is given by the imaginary part while $\Delta\Gamma_d$ is essentially given by the real part.
To leading order $\Gamma_{12}^d$ is the absorptive part of one loop diagrams and thus it is a product of tree-level amplitudes. The SM expression involves $u$ and $c$ intermediate quarks and reads \cite{Beneke:2003az,Ciuchini:2003ww}
\begin{equation}
\Gamma_{12}^d\propto F_2|\V{ud}\V{ub}|^2e^{-i2\gamma}+F_1|\V{ud}\V{ub}\V{cd}\V{cb}|e^{-i\gamma}+F_0|\V{cd}\V{cb}|^2~,
\label{Gamma12:01}
\end{equation} 
where $F_i$ are coefficients independent from CKM quantities (they will be specified when appropriate). Notice that \eq{Gamma12:01} is usually rewritten through the use of the unitarity relation $\Vc{ud}\V{ub}+\Vc{cd}\V{cb}+\Vc{td}\V{tb}=0$ in order to introduce $\Vc{td}\V{tb}$, as this combination of CKM elements also controls $M_{12}^d$; notwithstanding, we will keep \eq{Gamma12:01} because it does not require any assumption concerning the unitarity of the CKM matrix, although in the present framework $3\times 3$ unitarity is assumed. A first look at \eq{Gamma12:01} shows that there is indeed one term sensitive to the difference between $\gamma$ and $\gamma+\pi$, $F_1|\V{ud}\V{ub}\V{cd}\V{cb}|e^{-i\gamma}$, as its sign changes when $\gamma\to\gamma+\pi$: this should be kept in mind because it ultimately constitutes the origin of the usefulness of observables like $\Asld$ to distinguish SM-like solutions from NP ones in terms of $\gamma$ (see also \cite{Laplace:2002ik}). This sensitivity would also depend, obviously, on the numerical details: the coefficients $F_i$ and the ratio $|\V{cd}\V{cb}|/|\V{ud}\V{ub}|$; we will come to this aspect below. Let us consider now the denominator of $\Gamma_{12}^d/M_{12}^d$: the SM contribution to $M_{12}^d$ is dominated by the amplitude with intermediate top quarks,
\begin{equation}
[M_{12}^d]_{SM}=\frac{G_F^2M_W^2\fBd m_{B_d^0}\eta_{B}}{12\pi^2}(\V{tb}\Vc{td})^2 S_0(x_t)~.
\label{DMBd:01}
\end{equation}
The eventual presence of NP contributions is parametrised through
\begin{equation}
M_{12}^d=r_d^2 e^{-i2\phi_d}[M_{12}^d]_{SM}~.\label{DMBd:02}
\end{equation}
The crucial feature is that we already have \emph{experimental access} to $M_{12}^d$: first, as $\DMBd=2|M_{12}^d|$, the accurate measurement of the mass difference fixes quite well the modulus and second, the asymmetry $\AJPsi$ measures the phase, $\AJPsi=\sin(\arg M_{12}^d)=\sin(2\bar\beta)=\sin(2(\beta-\phi_d))$:
\begin{equation}
M_{12}^d=\frac{1}{2}\ \DMBd\ e^{i2\bar\beta}~.\label{DMBd:03}
\end{equation}
With equations (\ref{Gamma12:01}) and (\ref{DMBd:03}) we can write\footnote{The numerical factor $K_d\equiv\frac{10^{-4} G_F^2 M_W^2 \fBd m_{B_d^0}\eta_{B}S_0(x_t)}{6\pi^2}$ comes from the calculation of the coefficients $a$, $b$ and $c$ that enter the numerator in \eq{Gamma12M12d:01}; using equations (\ref{DMBd:01}), (\ref{DMBd:02}) and (\ref{DMBd:03}) it may be rewritten as $K_d=\frac{10^{-4}\DMBd}{r_d^2 |\V{tb}\Vc{td}|^2}$ but we keep the form of \eq{Gamma12M12d:01} to stress the fact that $M_{12}^d$, as expressed in \eq{DMBd:03}, is already a \emph{measured} quantity.}
\begin{multline}
\frac{\Gamma_{12}^d}{M_{12}^d}=2\frac{\Gamma_{12}^d}{\DMBd}e^{-i\arg(M_{12}^d)}=\frac{K_d}{\DMBd e^{i2\bar\beta}}\times\\((b+c-a)|\V{ud}\V{ub}|^2e^{-i2\gamma}+(a-2c)|\V{ud}\V{ub}\V{cd}\V{cb}|e^{-i\gamma}+c|\V{cd}\V{cb}|^2),\label{Gamma12M12d:01}
\end{multline}
where we explicitly show the coefficients $a=12.0\pm 2.4$, $b=0.2\pm 0.1$ and $c=-40.1\pm 15.8$ \cite{Beneke:2003az}.

\begin{figure}[h]
\begin{center}
\subfigure[$\Asld$ distribution\label{figure02a}]{\epsfig{file=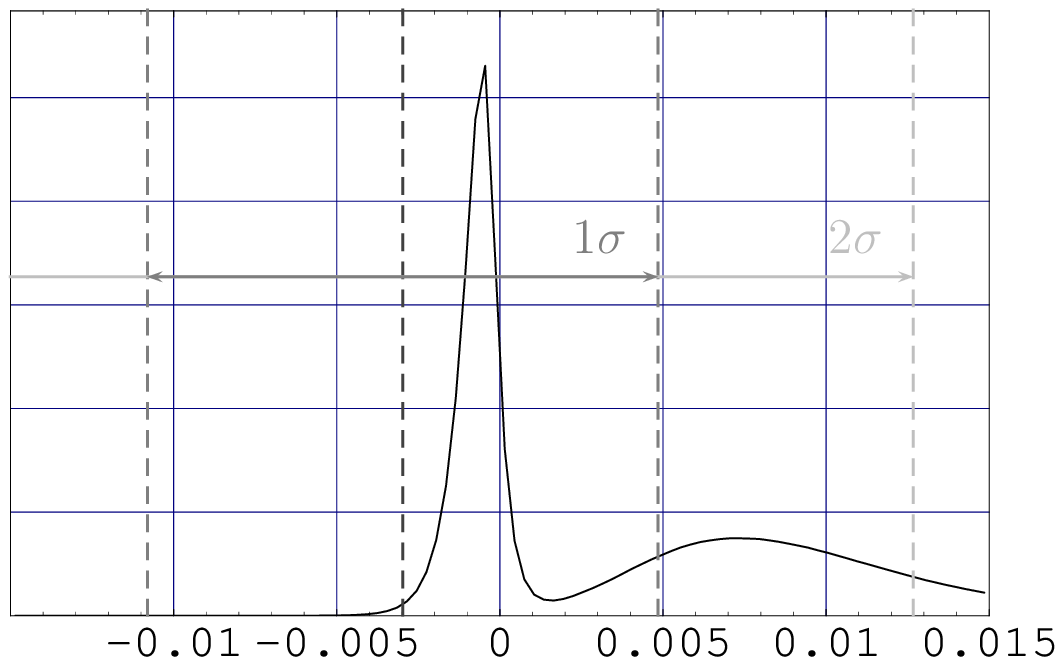,width=0.55\textwidth}}\quad \subfigure[$2\phi_d$ vs. $\Asld$ \label{figure02b}]{\epsfig{file=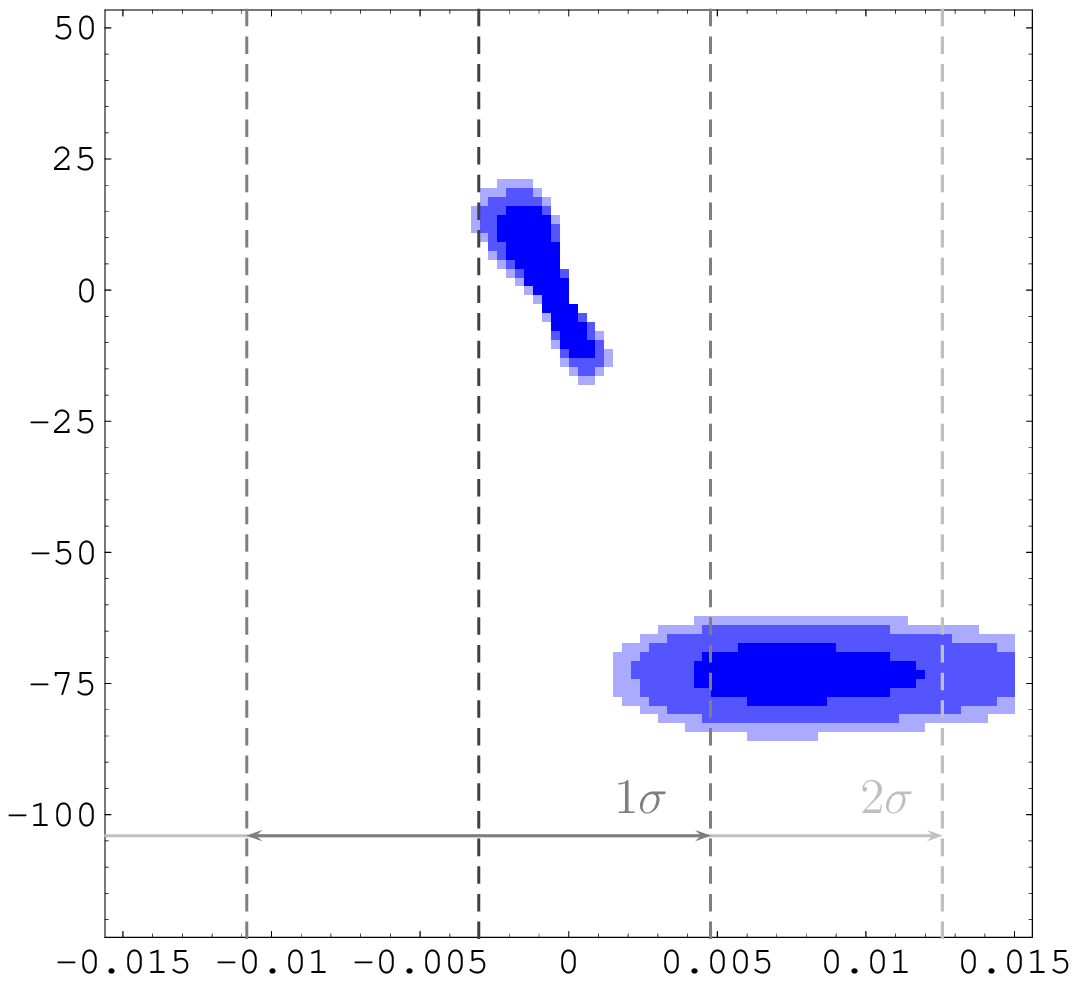,width=0.35\textwidth}}
\caption{Probability regions: $68\%$ dark, $90\%$ medium, $95\%$ light; the measured value of $\Asld$ is indicated through a dark dashed line, the uncertainty of this measurement is also shown in terms of $1\sigma$ and $2\sigma$ ranges.}\label{figure02}
\end{center}
\end{figure}

Equation (\ref{Gamma12M12d:01}) will be extremely useful to understand the r\^{o}le played by both the semileptonic asymmetry $\Asld$ and the width difference $\DGBd$. Let us recall that
\begin{equation}
\Asld=\im{\frac{\Gamma_{12}^d}{M_{12}^d}}\quad ; \quad \DGBd=-\DMBd\ \re{\frac{\Gamma_{12}^d}{M_{12}^d}}~, \label{ImReGamma12M12d}
\end{equation}
which leads to:
\begin{multline}
\Asld=\frac{K_d}{\DMBd}((b+c-a)|\V{ud}\V{ub}|^2\sin(2\bar\alpha)+\\(2c-a)|\V{ud}\V{ub}\V{cd}\V{cb}|\sin(2\bar\beta+\gamma)-c|\V{cd}\V{cb}|^2\underbrace{\sin(2\bar\beta)}_{\AJPsi}).\label{Asld:01}
\end{multline}
At this stage it is worth to make a rough evaluation of the size of the various terms contributing to $\Asld$. Using the fact that $|\V{cd}\V{cb}|\sim 3 |\V{ud}\V{ub}|$, $\sin(2\bar\alpha)\sim 0.35$, $\sin(2\bar\beta+\gamma)\sim \pm 1$ and $\AJPsi\sim 0.7$, together with the values of $a$, $b$, $c$, one obtains
\begin{eqnarray*}
(b+c-a)|\V{ud}\V{ub}|^2\sin(2\bar\alpha)&\sim& -20~ |\V{ud}\V{ub}|^2~, \\
(2c-a)|\V{ud}\V{ub}\V{cd}\V{cb}|\sin(2\bar\beta+\gamma)&\sim & \mp 275~|\V{ud}\V{ub}|^2~,\\
-c|\V{cd}\V{cb}|^2\AJPsi &\sim& 250~|\V{ud}\V{ub}|^2~.
\end{eqnarray*}
The first term is much smaller than the remaining ones; the interesting feature is that, depending on the sign of $\sin(2\bar\beta+\gamma)$, there may be a significant cancellation or not. For the SM-like solution,
\begin{multline}
(b+c-a)|\V{ud}\V{ub}|^2\sin(2\bar\alpha)+(2c-a)|\V{ud}\V{ub}\V{cd}\V{cb}|\sin(2\bar\beta+\gamma)\\ -c|\V{cd}\V{cb}|^2\sin(2\bar\beta)
\end{multline}
gives approximately $-45 |\V{ud}\V{ub}|^2$ while for the NP solution the corresponding result is $\sim 505 |\V{ud}\V{ub}|^2$.
This simple numerical exercise shows that the values of $\Asld$ corresponding to the SM-like solution will be negative and will have a size roughly $\mathcal O(10^{-3})$ while the values corresponding to the NP solution will be positive and will have a much larger size, $\mathcal O(10^{-2})$: the semileptonic asymmetry is, under those conditions, an obvious choice of observable sensitive to one or the other solution. Figure \ref{figure02a} displays the probability distribution of $\Asld$ and figure \ref{figure02b} the joint probability distribution of $\Asld$ and $2\phi_d$ obtained from a calculation using the basic set of constraints (\eq{basicset}) without including $\Asld$: the results of the previous estimate are confirmed. The figures also show the measured value of $\Asld$ and the corresponding uncertainty to underline the effectiveness that $\Asld$ may have to suppress the NP solution, even though it does not appear to be sufficient to really discard it.

\begin{figure}[h]
\begin{center}
\subfigure[$\DGBd/\Gamma_d$ distribution\label{figure03a}]{\epsfig{file=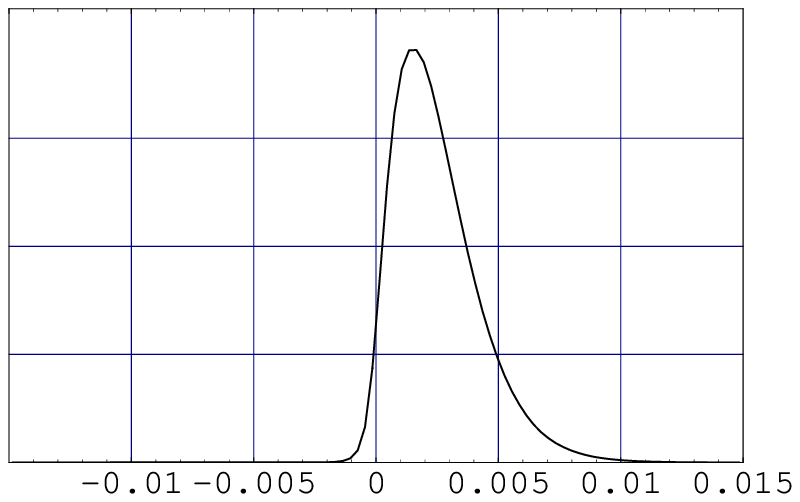,width=0.55\textwidth}}\quad \subfigure[$2\phi_d$ vs. $\DGBd/\Gamma_d$ \label{figure03b}]{\epsfig{file=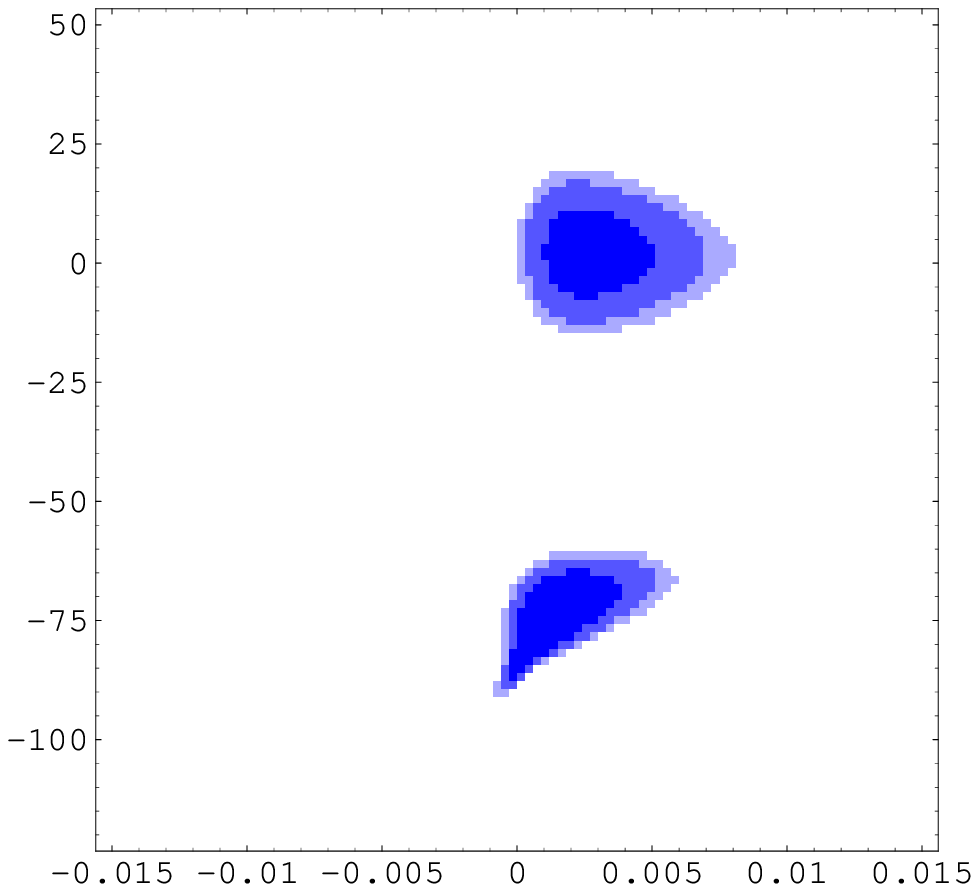,width=0.35\textwidth}}
\caption{Probability regions: $68\%$ dark, $90\%$ medium, $95\%$ light.}\label{figure03}
\end{center}
\end{figure}

We will now analyse how the real part of $\Gamma_{12}^d/M_{12}^d$, i.e. $\DGBd$, is going to be taken into account. Following equations (\ref{Gamma12M12d:01}) and (\ref{ImReGamma12M12d}) we have
\begin{multline}
\frac{\DGBd}{\Gamma_d}=-\frac{K_d}{\Gamma_d}((b+c-a)|\V{ud}\V{ub}|^2\cos(2\bar\alpha)\\ +(2c-a)|\V{ud}\V{ub}\V{cd}\V{cb}|\cos(2\bar\beta+\gamma)-c|\V{cd}\V{cb}|^2\cos(2\bar\beta)).\label{DGBd:01}
\end{multline}
In this case, the term depending on $2\bar\beta+\gamma$, the only one able to distinguish the SM-like from the NP solution, is simply suppressed because, with $2\bar\beta+\gamma\sim\pm\frac{\pi}{2}$, $\cos(2\bar\beta+\gamma)\sim 0$, and thus the values of $\DGBd$ computed for each case will not differ significantly, so this observable will not be very useful. In addition, the large uncertainty in the experimental determination of $\DGBd$ ($\DGBd/\Gamma_d=0.009\pm 0.037$, see table \ref{table:inputs}), compared to the calculated values, stresses the fact that this observable is not going to play any r\^{o}le in the present analyses. We show, however, the corresponding distributions in figure \ref{figure03} to illustrate this point.

\begin{figure}[h]
\begin{center}
\subfigure[Apex of the unitarity triangle $db$, $\im{-\frac{V_{ud}V_{ub}^\ast}{V_{cd}V_{cb}^\ast}}$ vs. $\re{-\frac{V_{ud}V_{ub}^\ast}{V_{cd}V_{cb}^\ast}}$\label{figure04a}]{\epsfig{file=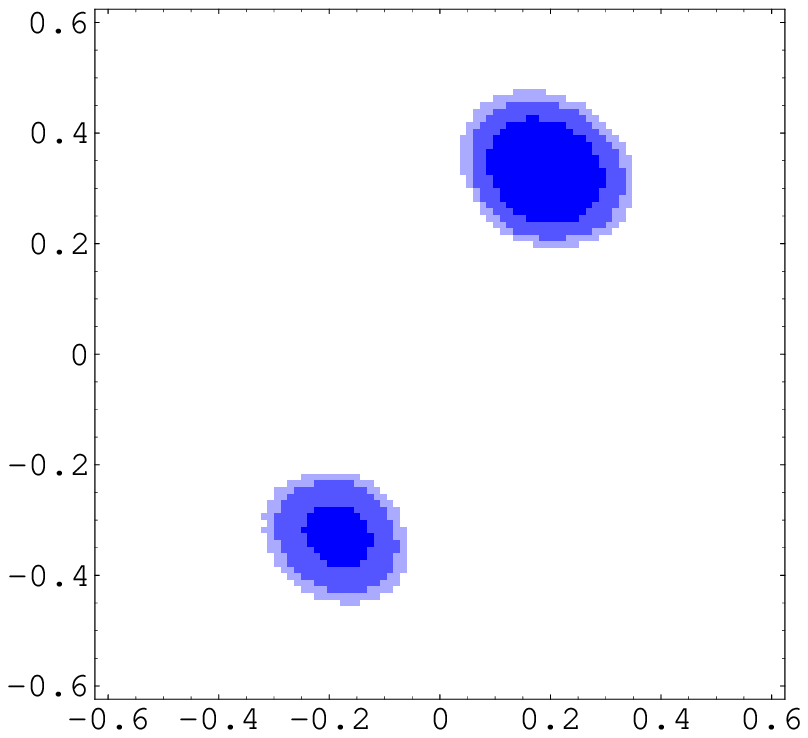,width=0.45\textwidth}}\quad \subfigure[$r_d^2$ vs. $2\phi_d$\label{figure04b}]{\epsfig{file=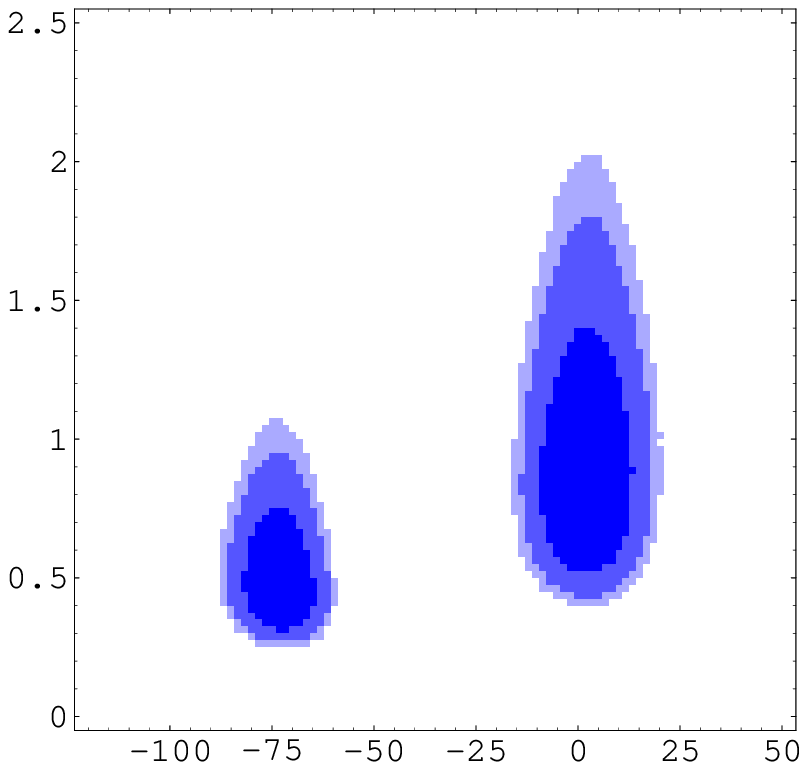,width=0.45\textwidth}}
\caption{Probability regions: $68\%$ dark, $90\%$ medium, $95\%$ light.}\label{figure04}
\end{center}
\end{figure}

Knowing the kind of impact that the use of $\Asld$ and $\DGBd$ will have in the determination of CKM and NP parameters, figure \ref{figure04} shows the numerical results obtained by including them among the constraints. The SM-like solution accumulates in this case $\sim$73.0\% of probability while the NP is at the $\sim$26.7\% level (the remaining two NP solutions accumulate 0.2\% and 0.1\% of the probability).

\subsubsection{New Physics tests in $\boldsymbol{\Gamma_{12}^d}$ with $\boldsymbol{\Asld}$}
Equation (\ref{Asld:01}) reflects the obvious fact that, being a genuinely CP-violating quantity, $\Asld$ may be written in terms of basic CKM rephasing invariant parameters that reflect CP violation. One can, however, make a different use of this equation. We recall that the only underlying assumption was, to consider New Physics is absent in tree level processes and therefore the absence of New Physics in the absorptive piece represented by $\Gamma_{12}^d$ (arbitrary NP has been allowed in $M_{12}^d$). The CKM moduli $|\V{ud}|$, $|\V{ub}|$, $|\V{cd}|$ and $|\V{cb}|$ are all independently measured through tree-level processes; the phases $2\bar\alpha$, $2\bar\beta+\gamma$ and $2\bar\beta$ are also measured in different decay channels. As a result, \eq{Asld:01} may be used as a test valid under rather general conditions, and thus provide an important test, once the experimental values reach a sufficient precision. In particular, non verification of \eq{Asld:01} could signal $3\times 3$ unitarity deviations and/or contamination  of penguins by NP loops (see for example \cite{Branco:1992uy}).

\section{NP analysis and $\boldsymbol{B_s^0}$--$\boldsymbol{\bar B_s^0}$ mixing \label{sec:BBs}}

\subsection{$\boldsymbol{\DMBs}$ and NP parameters}

\begin{figure}[h]
\begin{center}
\subfigure[$r_s^2$\label{figure05a}]{\epsfig{file=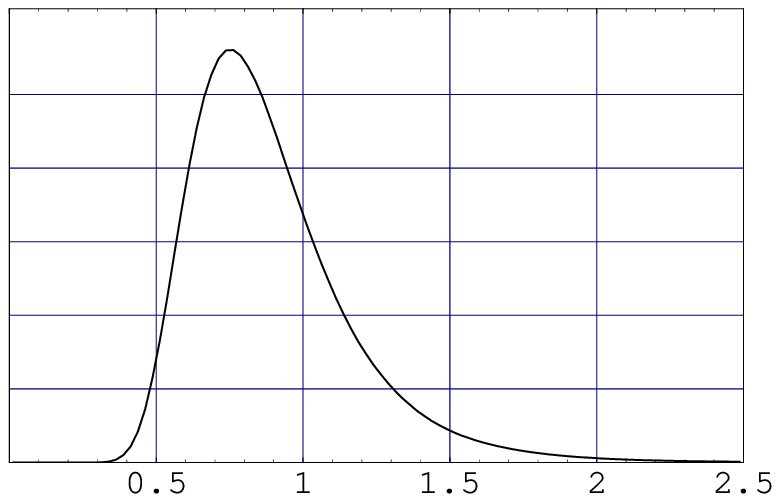,width=0.55\textwidth}}\quad \subfigure[$2\phi_d$ vs. $r_d^2/r_s^2$\label{figure05b}]{\epsfig{file=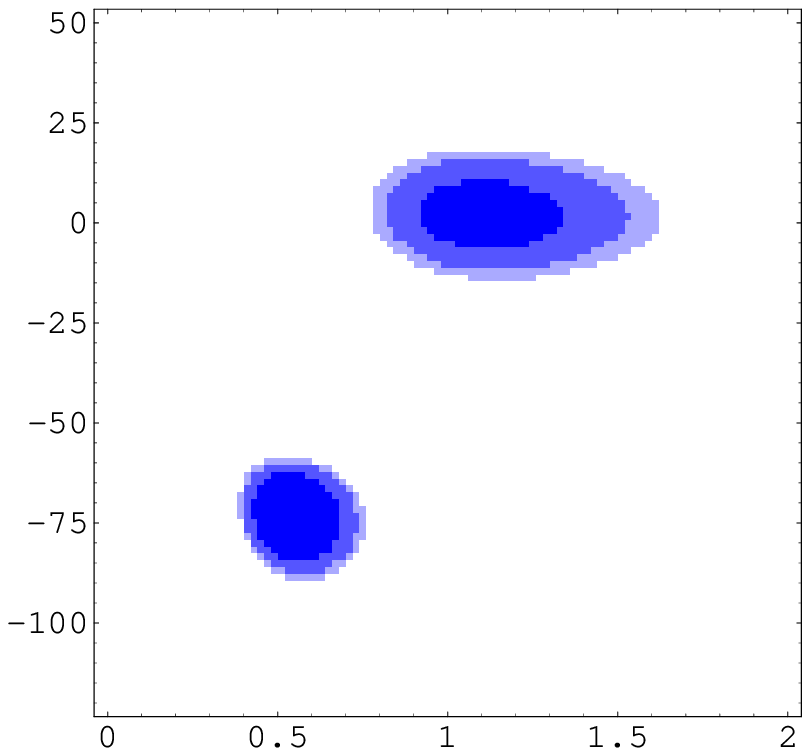,width=0.35\textwidth}}
\caption{Probability regions: $68\%$ dark, $90\%$ medium, $95\%$ light.}\label{figure05}
\end{center}
\end{figure}

Recent measurements at D0 and CDF \cite{Abazov:2006dm,Abulencia:2006mq} have provided the first measurements of $\DMBs$, going beyond the establishment of lower bounds. Using the usual general parametrisation of the \BBs~ mixing in the presence of NP contributions,
\[
M_{12}^s=r_s^2 e^{-i2\phi_s}[M_{12}^s]_{SM}~,
\]
it is straightforward to obtain the allowed range  for $r_s$. Taking into account that the measured value for $\DMBs$ is within the ranges predicted within the SM framework, it is expected that one of the solutions would be around $r_s\sim 1$. This is quite similar to the case of the SM-like solution with $r_d\sim 1$ and $\DMBd$ results for the \BBd~ mixing. At this stage, one could ask the question whether one expects to find a solution with $r_s$ differing significantly from $1$, in addition to the solution with $r_s\sim 1$. It can be readily seen that the answer to the above question is in the negative. Indeed from the unitarity triangles\footnote{We use the phase convention $\arg V=\left(\begin{smallmatrix}0& \chi^\prime & -\gamma\\ \pi & 0 & 0\\ -\beta & \pi+\chi & 0\end{smallmatrix}\right)$. Within $3\times 3$ unitarity $\chi^\prime\sim\mathcal O(\lambda^4)$, we will thus neglect $\chi^\prime$ in the following.} $db$ and $sb$, one obtains
\begin{eqnarray}
|\V{td}|=\frac{|\V{ud}\Vc{ub}+\V{cd}\Vc{cb}|}{|\V{tb}|}~,\label{unit:01}\\
|\V{ts}|=\frac{|\V{us}\Vc{ub}+\V{cs}\Vc{cb}|}{|\V{tb}|}~.\label{unit:02}
\end{eqnarray}
Changing $\gamma\to\gamma+\pi$ gives $\V{ub}\to -\V{ub}$. The numerator in \eq{unit:01} has two terms of order $\lambda^3$, a change of sign in one of them will produce a change in the result of order 1 and it is for this reason that $r_d$ can noticeably differ from 1 in the non SM-like solution. On the other hand, the numerator in \eq{unit:02} has two terms of different size, $\V{us}\Vc{ub}\sim \mathcal O(\lambda^4)$ and  $\V{cs}\Vc{cb}\sim \mathcal O(\lambda^2)$, a change of sign in the first one will only imply a small change in the value of the numerator and thus no significant deviation\footnote{Once again, uncertainties in several parameters will produce wider ranges than simple arguments make us expect; in fact, this may be used the other way around, precise determinations of $\DMBd$ and $\DMBs$ help to constrain some of those hadronic parameters because they are in fact the main source of uncertainty in the calculations.} can be expected for $r_s$. 
Notice, however, that the maximum of the $r_s$ distribution in figure \ref{figure05a} is not exactly at $r_s=1$; there is reason for this: the predicted SM values of $\DMBs$, because of the dominant uncertainties coming from hadronic parameters, span a range roughly going from 15 ps$^{-1}$ to 30 ps$^{-1}$. Even if the measurement is safely installed within this range, it is in the low-values region and thus $r_s^2$, being nothing else but the ratio of the measured $\DMBs$ and what would be the SM prediction of it, will have a tendency to be smaller than $1$. This feature is completely manifest in figure \ref{figure05a}.
In figure \ref{figure05} we show the probability distribution of $r_s^2$ and the joint $(r_d^2/r_s^2,2\phi_d)$ probability distribution obtained with the basic set of observables in \eq{basicset} and $\DMBs$. $\Asld$ has not been used to obtain these distributions in order to analyse separately the effect of the different observables. Even though the set of constraints used in figure \ref{figure05} is still far from spanning the whole set used in section \ref{sec:Complete}, one can draw a few interesting conclusions. NP in both \BBd~ and \BBs~ is described by two definite ranges of values of $r_d^2/r_s^2$, either close to 1 or close to 0.6 because, as mentioned before, $r_s\sim 1$ and $r_d$ may have two allowed regions depending on the value of $\gamma$: New Physics models of the Minimal Flavour Violating (MFV) type (see \cite{Blanke:2006ig} and references therein), which have received significant attention in the literature, may not trivially produce $r_d/r_s\neq 1$. In this sense, within this general framework, NP contributions to the studied mixings should be either small when compared to the SM ones (the repeated successes of the SM strengthen this trend) or significant and involving a richer flavour structure, including new sources of CP violation.

\begin{figure}[h]
\begin{center}
\subfigure[$r_s^2$ vs. $2\phi_s$\label{figure06a}]{\epsfig{file=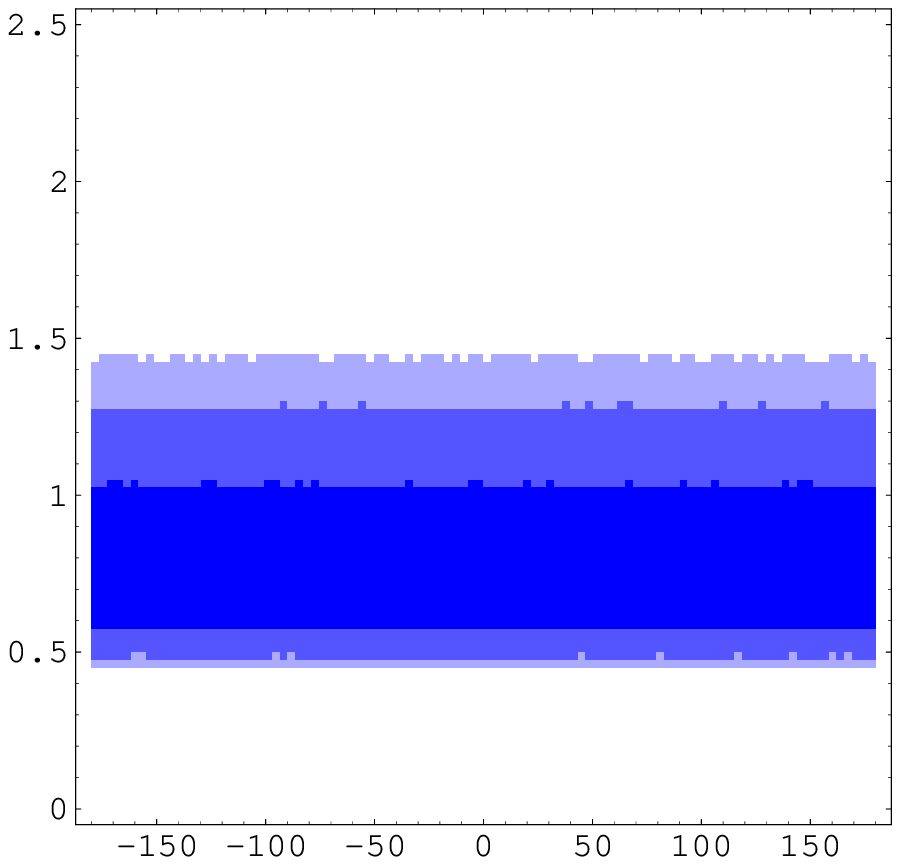,width=0.45\textwidth}}\quad \subfigure[$h_s$ vs. $\sigma_s$\label{figure06b}]{\epsfig{file=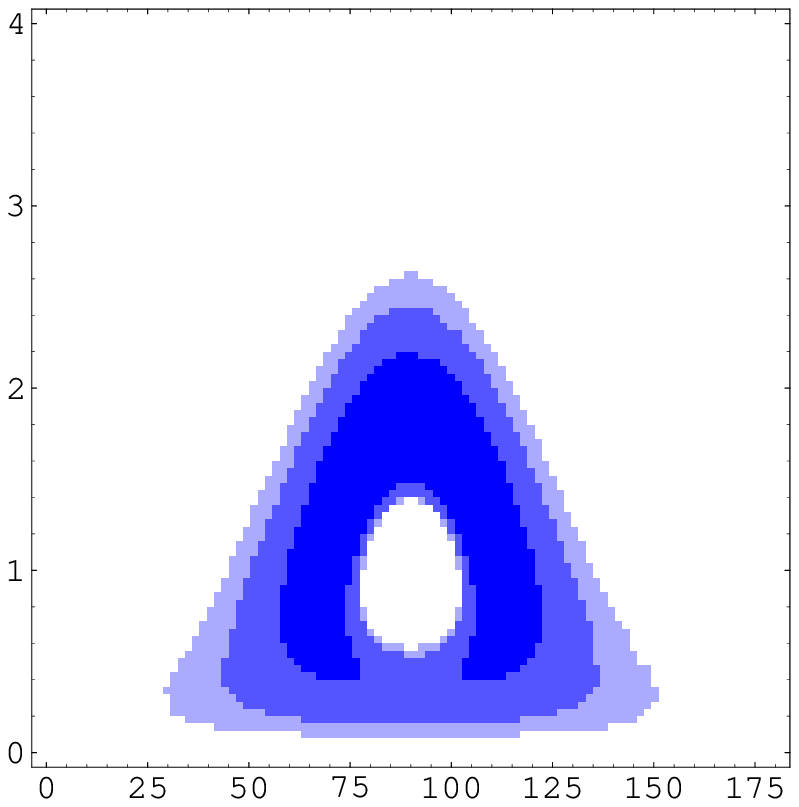,width=0.45\textwidth}}
\caption{Probability regions: $68\%$ dark, $90\%$ medium, $95\%$ light.}\label{figure06}
\end{center}
\end{figure}

At present, no observable sensitive to the phase of $M_{12}^s$ has been measured so $2\phi_s$ is completely free; if we where to use a different parametrization for NP in $M_{12}^s$, as is done by some authors \cite{Ligeti:2006pm,Ball:2006xx,Baek:2006bv}, the relation $r_s^2e^{-i2\phi_s}=1+h_se^{i2\sigma_s}$ would produce probability distributions in parameter space looking quite different, which is illustrated in figure \ref{figure06}. Notice that a simple result like $r_s^2=0.84\pm 0.26$ \cite{Isidori:2006pk} has no direct translation to $h_s$ and $\sigma_s$. One should keep in mind that, taking into account the small number of observables now available to explore NP in $B_s$ mesons and thus to constrain the NP parameter space, a simultaneous look to both parameterisations is required. One concludes from figure \ref{figure06} that $h_s$ has an upper bound $h_s\sim 2$ for $2\sigma_s\sim \pm\pi$ which means that the combination $1+h_se^{i2\sigma_s}$(times the SM prediction) is reproducing the SM-compatible experimental result by setting the NP contribution to minus twice the SM prediction, thus giving an overall result which is just the SM prediction with the sign changed. Note that $h_s$ represents the ratio of the NP contributions and the SM ones.

\subsection{The r\^{o}le of $\boldsymbol{\Gamma_{12}^s/ M_{12}^s}$: $\boldsymbol{\Ads}$ and $\boldsymbol{\DGBsCP}$}

The next worth studying observables that could enter our analyses are related to $\Gamma_{12}^s/ M_{12}^s$; the procedure to understand their impact on the results will be quite similar to the one followed in section \ref{sec:BBd}, and the expressions appropriate to the \BBs~ case can be easily obtained from the \BBd~ ones (\eq{Gamma12M12d:01}) through the changes $m_{B_d^0}\to m_{B_s^0}$, $\fBd\to \fBs$, to get\footnote{Instead of $\fBd$ and $\fBs$ we will rather use in our calculations $\fBs$ and $\xi^2\equiv\frac{\fBs}{\fBd}$, as lattice QCD results for $\xi$ benefit from cancellations that reduce the uncertainty.} $K_d\to K_s$, and the simple substitutions 
\begin{equation}
|\V{ud}|\to |\V{us}|e^{i\chi\prime},\quad |\V{cd}|\to -|\V{cs}|,\quad |\V{td}|e^{-i\beta}\to |\V{ts}|e^{i\chi}~.\label{BdtoBs}
\end{equation}
One then obtains:
\begin{multline}
\frac{\Gamma_{12}^s}{M_{12}^s}=\frac{K_s}{\DMBs e^{-i2\bar\chi}}\times\\((b+c-a)|\V{us}\V{ub}|^2e^{-i2\gamma}+(2c-a)|\V{us}\V{ub}\V{cs}\V{cb}|e^{-i\gamma}+c|\V{cs}\V{cb}|^2)~.\label{Gamma12M12s:01}
\end{multline}
\eq{Gamma12M12s:01} shows two main differences with respect to the \BBd~ case: first, the phase of $M_{12}^s$, $-2\bar\chi=-2(\chi+\phi_s)$, is completely arbitrary, and second, $\Gamma_{12}^s$ does not have three contributions of -- roughly -- the same size, there is instead a definite hierarchy: $|\V{us}\V{ub}|^2\sim \mathcal O(\lambda^8)$, $|\V{us}\V{ub}\V{cs}\V{cb}|\sim \mathcal O(\lambda^6)$ and $|\V{cs}\V{cb}|^2\sim \mathcal O(\lambda^4)$.

As in section \ref{sec:BBd}, we can calculate the semileptonic asymmetry $\Asls$ and the width difference $\DGBs$,
\begin{equation}
\Asls=\im{\frac{\Gamma_{12}^s}{M_{12}^s}}\quad ; \quad \DGBs=-\DMBs\ \re{\frac{\Gamma_{12}^s}{M_{12}^s}} \label{ImReGamma12M12s}~.
\end{equation}
Those quantities are not directly measured and they will require some additional discussion in the following paragraphs.

\begin{figure}[h]
\begin{center}
\subfigure[$2\phi_d$ vs. $\Asls$\label{figure07a}]{\epsfig{file=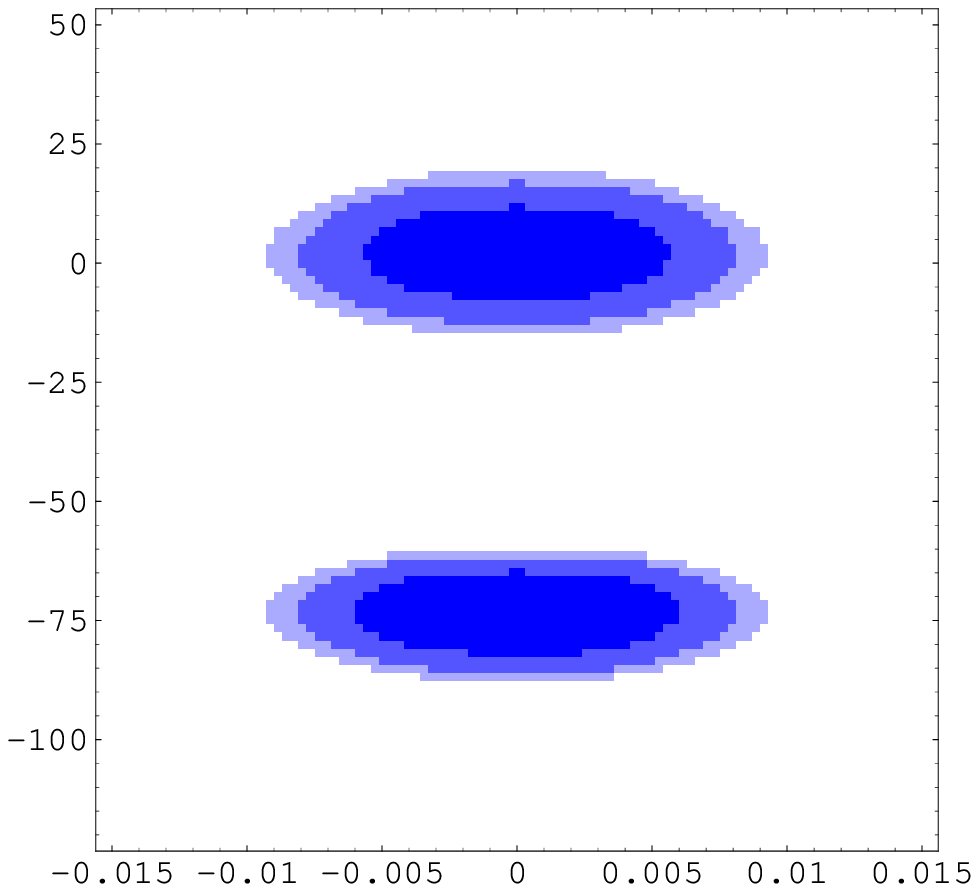,width=0.45\textwidth}}\quad \subfigure[$2\phi_s$ vs. $\Asls$ \label{figure07b}]{\epsfig{file=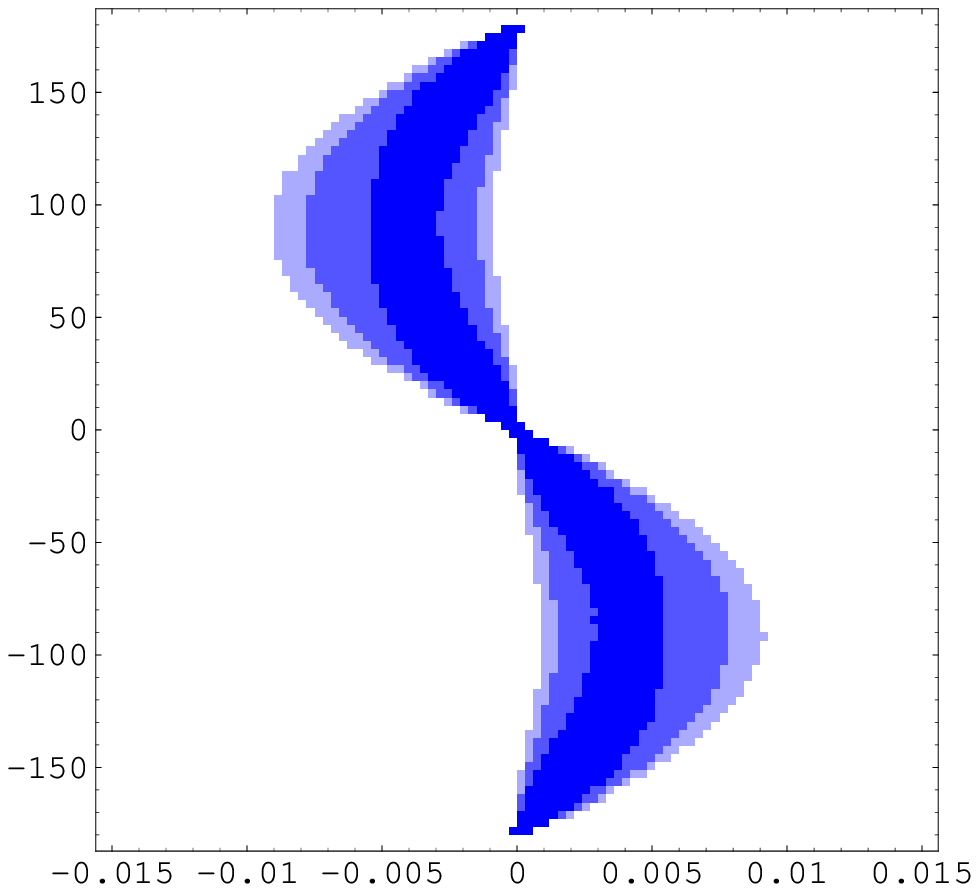,width=0.45\textwidth}}
\caption{Probability regions: $68\%$ dark, $90\%$ medium, $95\%$ light.}\label{figure07}
\end{center}
\end{figure}

\begin{figure}[h]
\begin{center}
\subfigure[$2\phi_d$ vs. $\Ads$\label{figure08a}]{\epsfig{file=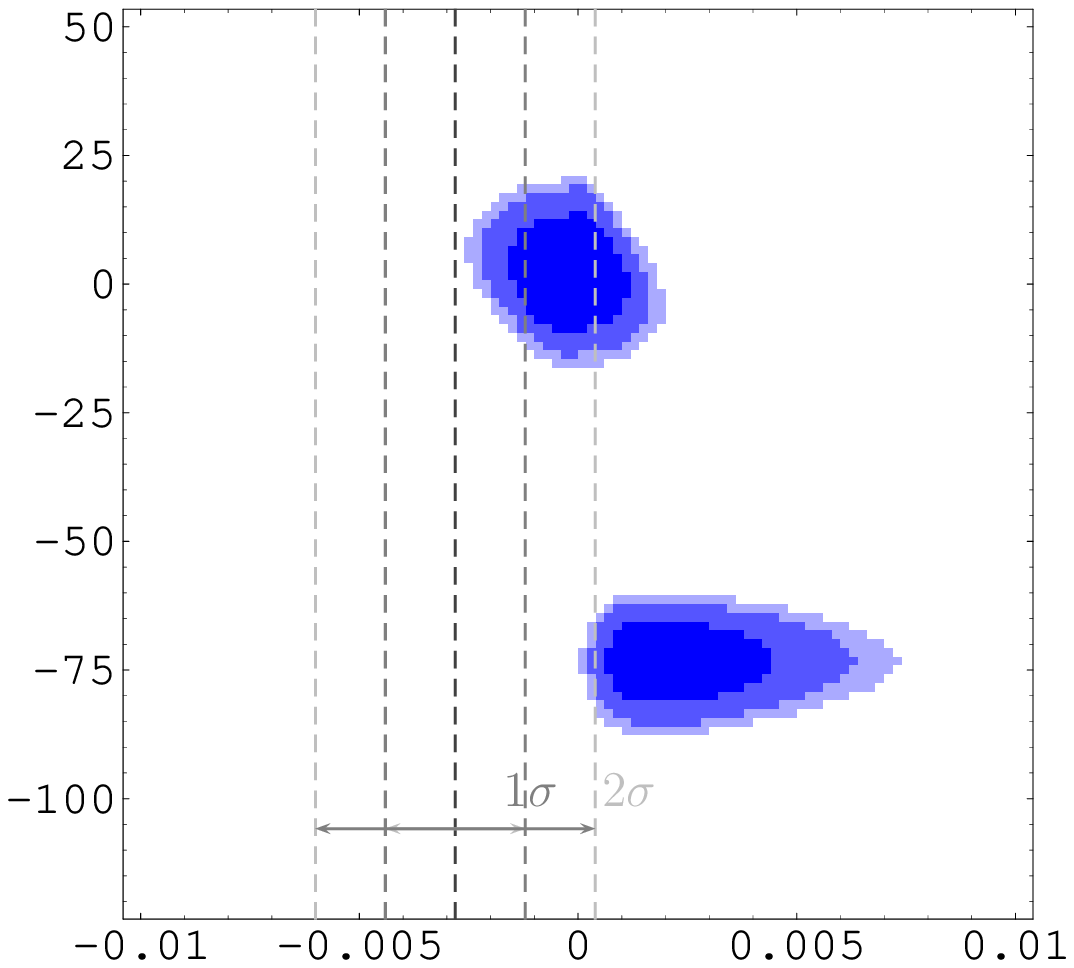,width=0.45\textwidth}}\quad \subfigure[$2\phi_s$ vs. $\Ads$\label{figure08b}]{\epsfig{file=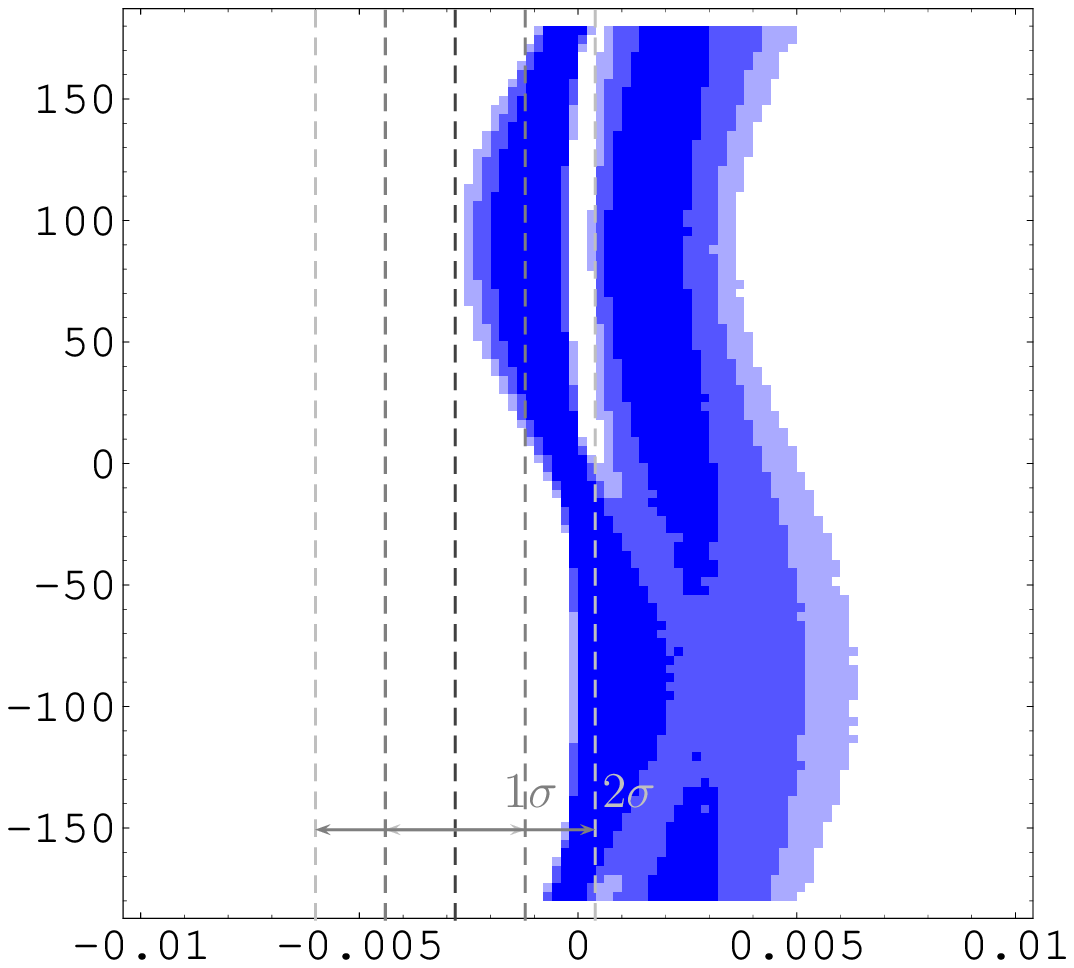,width=0.45\textwidth}}
\caption{Probability regions: $68\%$ dark, $90\%$ medium, $95\%$ light.}\label{figure08}
\end{center}
\end{figure}

Concerning the semileptonic asymmetry, it is not directly accessible in collider experiments like D0 and CDF, both $B_d$ and $B_s$ species are produced and thus asymmetries of this kind involve both individual asymmetries. We need, however, to calculate $\Asls$:

\begin{multline}
\Asls=\frac{K_s}{\DMBs}((b+c-a)|\V{us}\V{ub}|^2\sin(2[\bar\chi-\gamma])+\\ (2c-a)|\V{us}\V{ub}\V{cs}\V{cb}|\sin(2\bar\chi-\gamma)+c|\V{cs}\V{cb}|^2\sin(2\bar\chi)).\label{Asls:01}
\end{multline}
Taking into account that in the SM $2\bar\chi=2\chi\sim\lambda^2$, as $\bar\chi=\chi+\phi_s$ is a free parameter, the hierarchy among the CKM matrix elements in \eq{Asls:01} implies that the last term is the dominant one, over almost all the parameter space; this term is independent of $\gamma$ and therefore independent of $\phi_d$.
This is illustrated in figure \ref{figure07a}, where the joint probability distribution of $(2\phi_d,\Asls)$ is computed by taking only into account the basic set of constraints in \eq{basicset}. It is easily seen that the range of variation of $\Asls$ does not differ significantly from one region to the other, confirming our na\"{\i}ve guess. In opposition to the insensitivity to $\gamma$ and thus to $\phi_d$, the sinusoidal dependence of $\Asls$ on $2\bar\chi$ does imply an important sensitivity to $\phi_s$: $\Asls$ would be the first observable considered in this analysis that can help us to gain information on $\phi_s$, as shown in figure \ref{figure07b}. Nevertheless, as emphasized in \cite{Grossman:2006ce}, the measured quantity \cite{D0ASL} is not $\Asls$ but the dimuon charge asymmetry $\Ads$, given by:
\begin{equation}
\Ads=\frac{1}{4f}\left(\Asld+\frac{f_sZ_s}{f_dZ_d}\Asls\right)~,\label{Ads}
\end{equation}
where $f=0.814\pm 0.105$, $f_d=0.4$ and $f_s=0.1$ are the respective production fractions and 
\[
Z_q=\frac{1}{1-\left(\frac{\Delta\Gamma_q}{2\Gamma_q}\right)^2}-\frac{1}{1+\left(\frac{\Delta M_{B_q}}{\Gamma_q}\right)^2}~.
\]
Considering the previous discussion on $\Asls$, the importance of $\Ads$ is twofold: first, as it includes $\Asld$, which, as analysed in section \ref{sec:BBd}, is really sensitive to the presence of NP in the \BBd~ mixing, it may give information on $\phi_d$, and second, as it includes $\Asls$, it may also give information on $\phi_s$ . Figure \ref{figure08}, showing the joint $(2\phi_d,\Ads)$ and $(2\phi_s,\Ads)$ probability distributions together with $1\sigma$ and $2\sigma$ measured ranges of $\Ads$, illustrates quite well this point. The important r\^ole played by the experimental measurement of $\Ads$ in order to suppress the NP solution with $\gamma\sim -115^\circ$ is manifest in figure \ref{figure08a}. From figure \ref{figure08b} one can also learn that $\Ads$ will also favour values of $2\phi_s\sim 90^\circ$.

\begin{figure}[h]
\begin{center}
\subfigure[$2\phi_d$ vs. $\DGBsCP$\label{figure09a}]{\epsfig{file=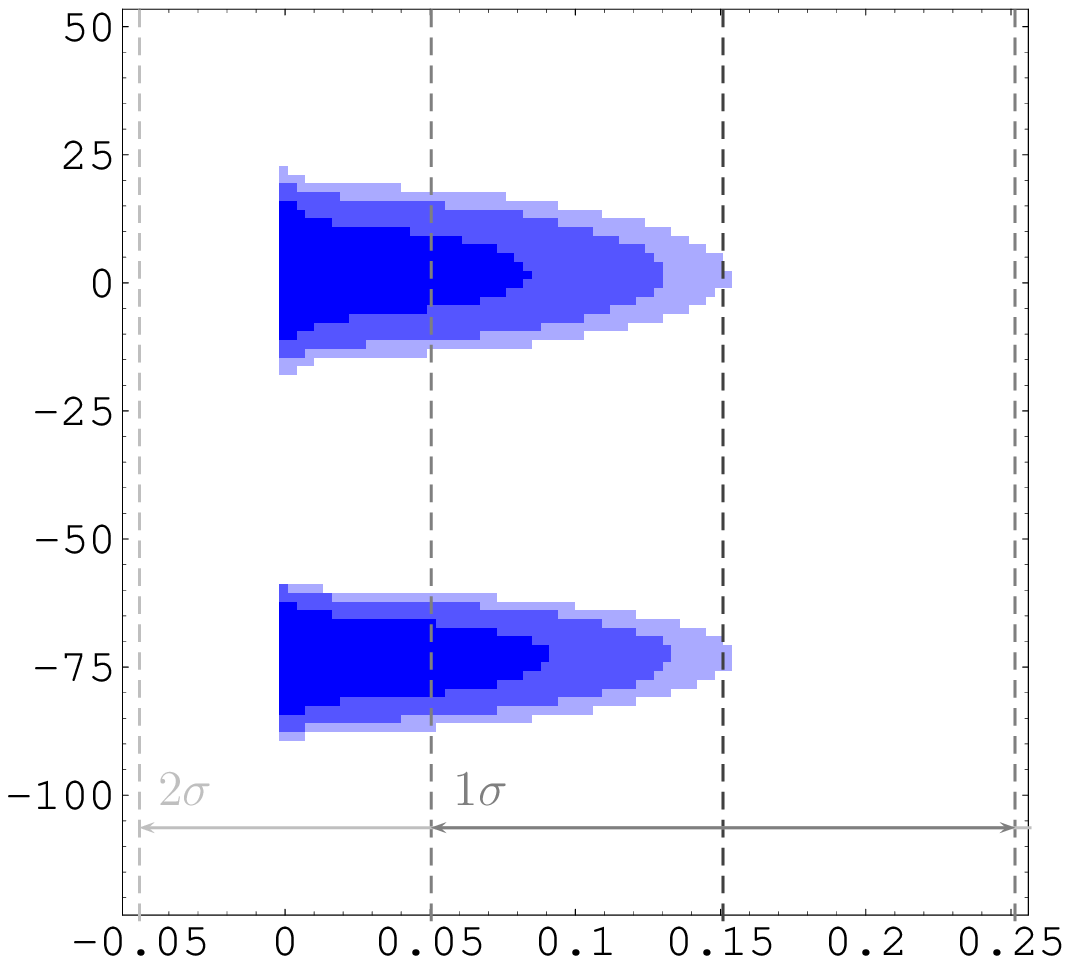,width=0.45\textwidth}}\quad \subfigure[$2\phi_s$ vs. $\DGBsCP$\label{figure09b}]{\epsfig{file=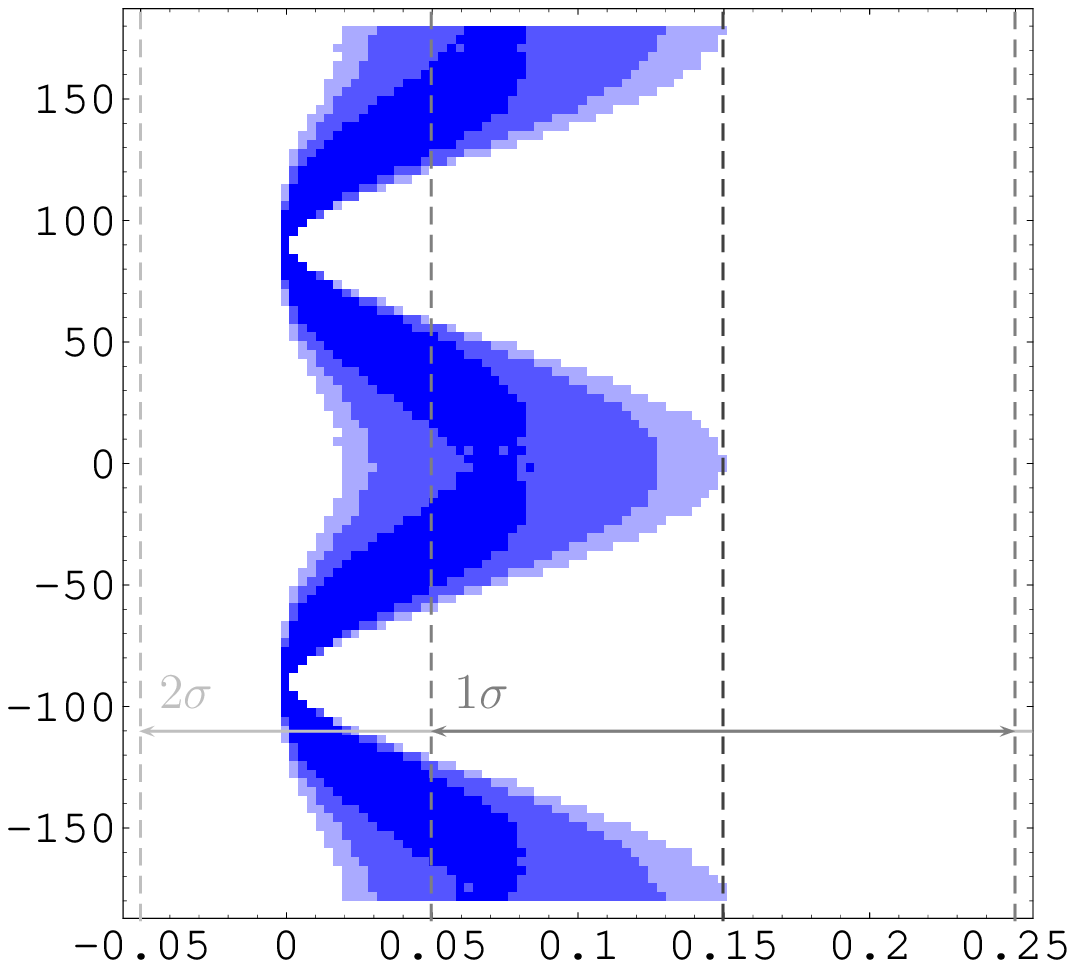,width=0.45\textwidth}}
\caption{Probability regions: $68\%$ dark, $90\%$ medium, $95\%$ light.}\label{figure09}
\end{center}
\end{figure}

Finally, using \eqs{Gamma12M12s:01} and (\ref{ImReGamma12M12s}) one obtains for the width difference:
\begin{multline}
\DGBs=-K_s((b+c-a)|\V{us}\V{ub}|^2\cos(2[\bar\chi-\gamma])+\\ (2c-a)|\V{us}\V{ub}\V{cs}\V{cb}|\cos(2\bar\chi-\gamma)+c|\V{cs}\V{cb}|^2\cos(2\bar\chi)).\label{DGBs:01}
\end{multline}
As in \eq{Asls:01}, the dominant term in \eq{DGBs:01} is the last one for almost any value of $2\bar\chi$. D0 and CDF do not measure, however, $\DGBs$ but $\DGBsCP$ \cite{D0DGBs,CDFDGBs}, the width difference between CP eigenstates. These quantities are related through \cite{Grossman:1996er,Dunietz:2000cr}
\begin{equation}
\DGBsCP=\DGBs \cos(2\bar\chi)~.\label{DGBsDGBsCP}
\end{equation}
This additional $\cos(2\bar\chi)$ factor does not change, however, what was stated a few lines above, the dominant term is the same third one; it will, additionally, force $\DGBsCP$ to be positive for almost all the parameter space. As it was the case for $\Ads$, $\DGBsCP$ illustrates an important difference between the information we can obtain on NP contributions to \BBd~ and \BBs~ mixings: despite significant NP in \BBd, associated with the $\gamma+\pi$ solution (with $r_d\neq 1$ and $\phi_d\neq 0$), it will not have a related sizable impact on NP determination in \BBs. On the other hand, taking into account the strong dependence of $\DGBsCP$ on $2\bar\chi$, up to the experimental uncertainty reached in its measurement, it may be interestingly sensitive to NP in \BBs~ without changing the picture in \BBd.
Figure \ref{figure09} shows the probability distributions of $(2\phi_d,\DGBsCP)$ and $(2\phi_s,\DGBsCP)$ obtained with the sole imposition of the basic set of observables. The insensitivity to $2\phi_d$ is obvious in figure \ref{figure09a}; the sensitivity to $2\phi_s$ is easily understood with figure \ref{figure09b}. The experimental result is $\DGBsCP=(0.15\pm 0.11)$ ps$^{-1}$, which gives a central value just on the edge of the predicted range, this fact will imply that values of $\phi_s$ which make the calculated $\DGBsCP$ as large as possible will be favoured; those are clearly the ones that render $\cos^2(2\bar\chi)$ as large as possible, that is $2\bar\chi\sim 2\phi_s\sim 0,\pm\pi$. This feature, together with the implications of figure \ref{figure08b}, will show up in the complete analysis of section \ref{sec:Complete}.

\subsection{$\boldsymbol{\Ads}$, $\boldsymbol{\DGBsCP}$ and $\boldsymbol{\AJPsiPhi}$}
\begin{figure}[h]
\begin{center}
\subfigure[$\AJPsiPhi$ vs. $\Ads$\label{figure10a}]{\epsfig{file=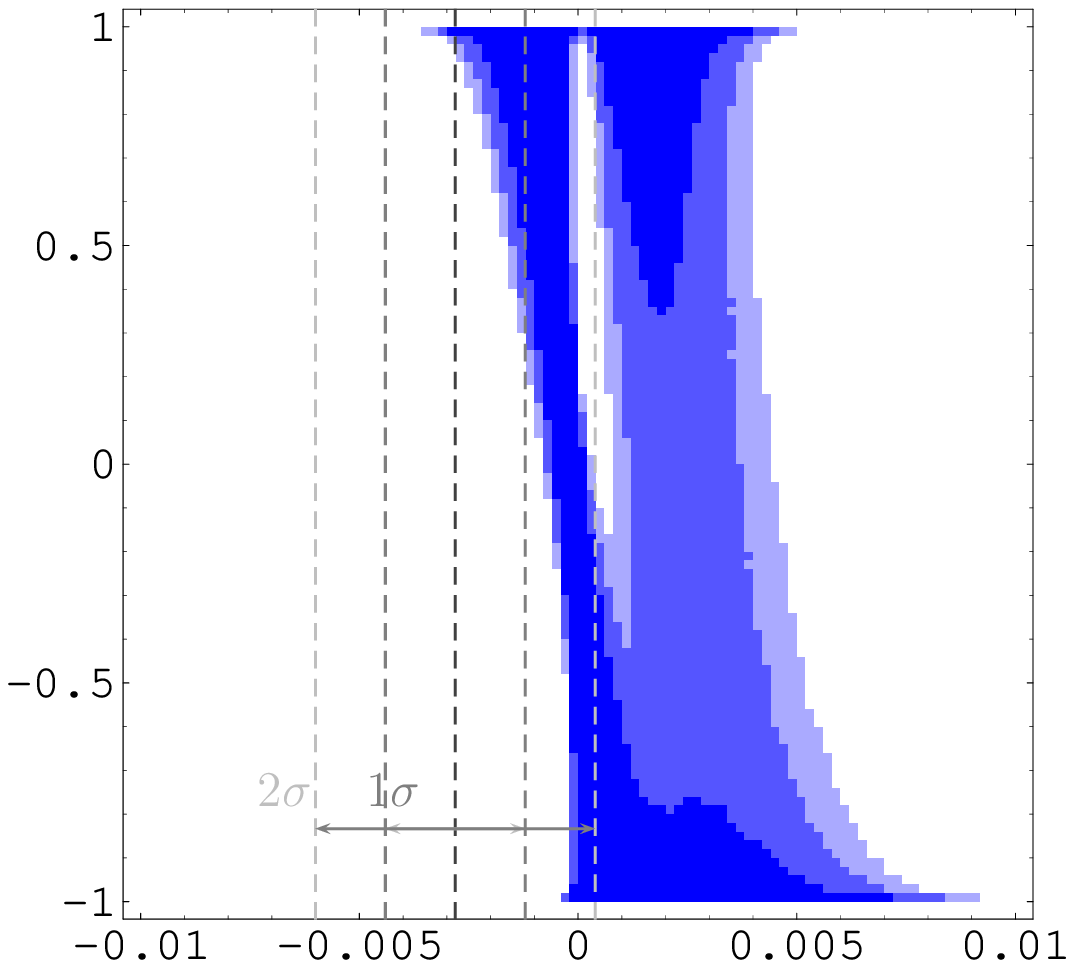,width=0.45\textwidth}}\quad \subfigure[$\AJPsiPhi$ vs. $\DGBsCP$\label{figure10b}]{\epsfig{file=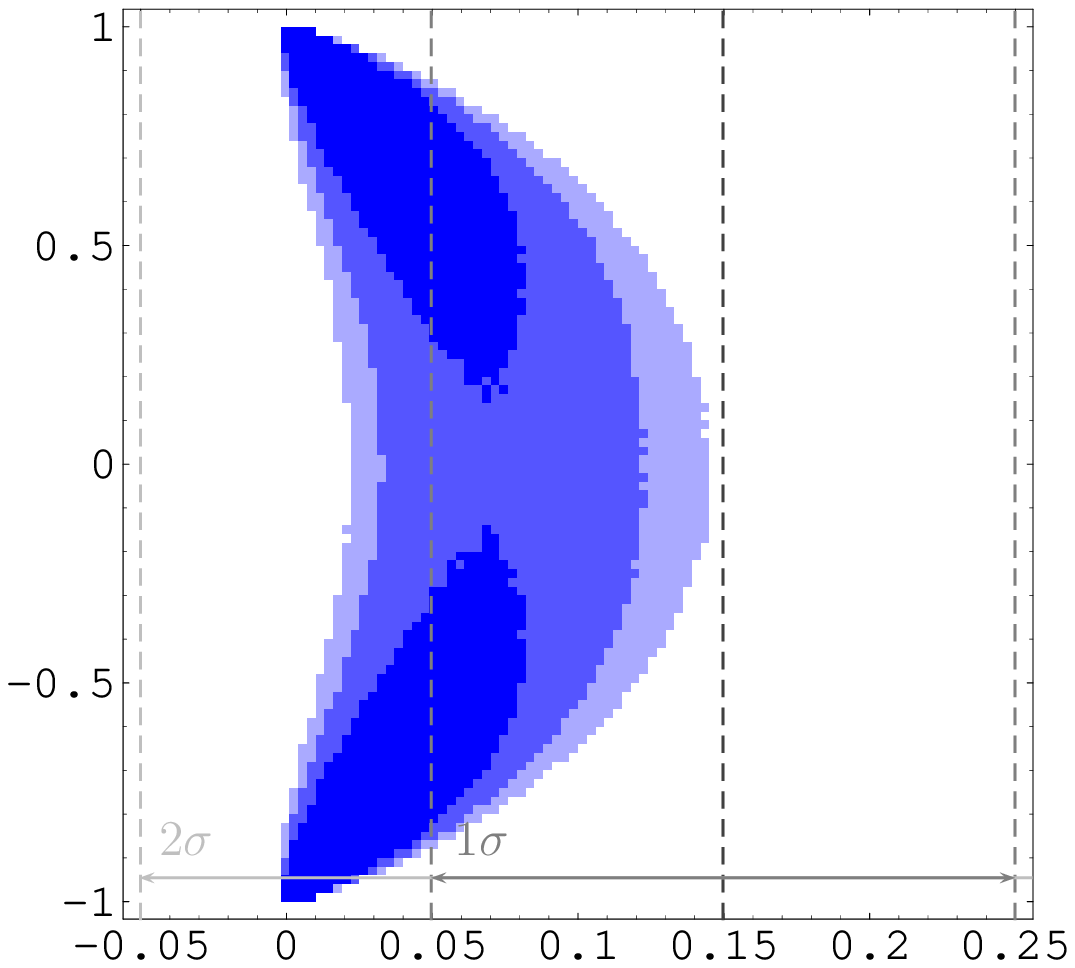,width=0.45\textwidth}}
\caption{Probability regions: $68\%$ dark, $90\%$ medium, $95\%$ light.}\label{figure10}
\end{center}
\end{figure}

In the previous subsection we have analysed $\Ads$ and $\DGBsCP$ giving a detailed account of their dependence on $\phi_s$; while in the \BBd\ system we have direct access to phases like $2\bar\beta$, $2\bar\beta+\gamma$, $\gamma$ or $\bar\alpha$ through different channels, we still have to wait before any measurement provides a direct constraint on $\phi_s$ through the $\bar\chi$ dependence of the \BBs\ mixing. The most interesting and obvious candidate is the time dependent asymmetry in $B_s^0,\bar B_s^0\to J/\Psi\Phi$ with a definite CP final state:
\begin{equation}
\AJPsiPhi=\sin(2\bar\chi)~,
\end{equation}
which could provide direct information on $\phi_s$ \cite{Blanke:2006ig,Ligeti:2006pm,Grossman:2006ce}. It is straightforward to rewrite \eq{Asls:01} including $\AJPsiPhi$:
\begin{multline}
\Asls\simeq \frac{K_s}{\DMBs}[c|\V{cs}\V{cb}|^2\AJPsiPhi+\\ (2c-a)|\V{cs}\V{cb}\V{us}\V{ub}|\{\AJPsiPhi\cos\gamma\pm\sqrt{1-\AJPsiPhi^2}\sin\gamma\}]~,
\end{multline}
where we neglect the term proportional to $|\V{us}\V{ub}|^2$; the dominant contribution is just linear in $\AJPsiPhi$. In figure \ref{figure10a} we show the joint $(\Ads,\AJPsiPhi)$ probability distribution obtained with the usual basic set of constraints in \eq{basicset}; the presence of \emph{two} linear branches is easily understood in terms of the \emph{two} different ranges of predicted $\Asld$ entering $\Ads$; the joint $(\Asls,\AJPsiPhi)$ probability distribution does not have two distinct branches,which reflects the insensitivity of $\Asls$ to the difference between $\gamma\sim 65^\circ$ and $\gamma\sim -115^\circ$ (see for example $\Asls$ vs. $\AJPsiPhi$ in reference \cite{Ligeti:2006pm}). It is also straightforward to rewrite \eq{DGBs:01} in terms of $\AJPsiPhi$:
\begin{multline}
\DGBsCP\simeq K_s [c|\V{cs}\V{cb}|^2(1-\AJPsiPhi^2)+\\ (2c-a)|\V{cs}\V{cb}\V{us}\V{ub}|\{(1-\AJPsiPhi^2)\cos\gamma\pm \AJPsiPhi\sqrt{1-\AJPsiPhi^2}\sin\gamma\}]~,
\end{multline}
where we have also neglected the $|\V{us}\V{ub}|^2$ term. The additional $\cos(2\bar\chi)$ factor in \eq{DGBsDGBsCP} produces in this case a nonlinear $\DGBsCP$ vs. $\AJPsiPhi$ dependence; this is illustrated with the joint $(\DGBsCP,\AJPsiPhi)$ probability distribution of figure \ref{figure10b}, also obtained with the basic set of constraints (\eq{basicset}). The \emph{predicted} probability distribution of $2\bar\chi$ making use of the full set of available constraints will be shown in the next section.

\section{Complete analysis \label{sec:Complete}}

\begin{figure}[h]
\begin{center}
\subfigure[Apex of the unitarity triangle $db$, $\im{-\frac{V_{ud}V_{ub}^\ast}{V_{cd}V_{cb}^\ast}}$ vs. $\re{-\frac{V_{ud}V_{ub}^\ast}{V_{cd}V_{cb}^\ast}}$\label{figure11a}]{\epsfig{file=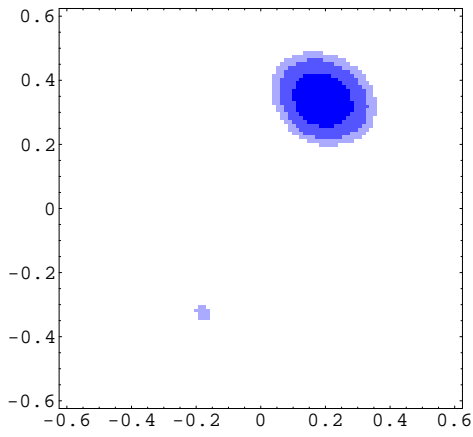,width=0.45\textwidth}}\quad \subfigure[$2\phi_d$ vs. $r_d^2/r_s^2$\label{figure11b}]{\epsfig{file=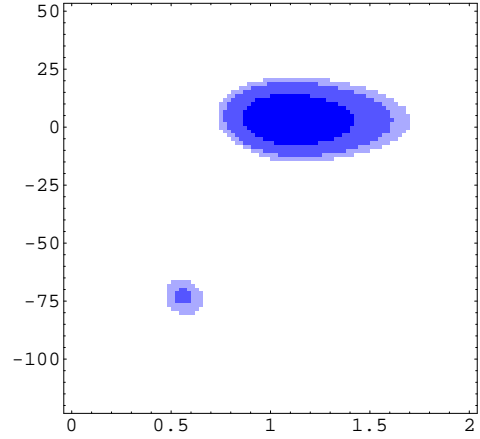,width=0.45\textwidth}}\\
\subfigure[$r_d^2$ vs. $2\phi_d$\label{figure11c}]{\epsfig{file=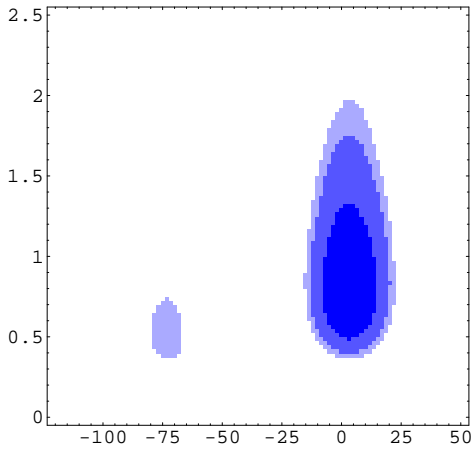,width=0.45\textwidth}}\quad \subfigure[$r_s^2$ vs. $2\phi_s$\label{figure11d}]{\epsfig{file=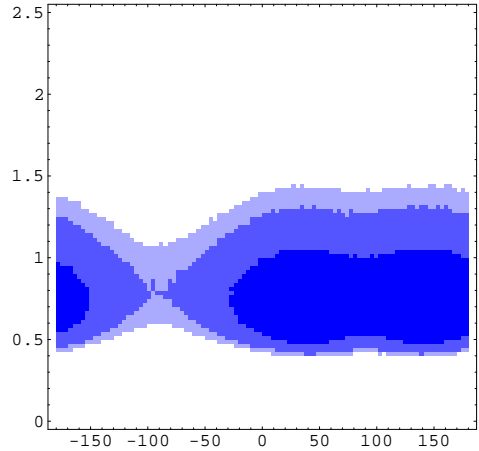,width=0.45\textwidth}}
\caption{Probability regions: $68\%$ dark, $90\%$ medium, $95\%$ light.}\label{figure11}
\end{center}
\end{figure}

In the previous sections we have addressed the study of several observables sensitive to \BBd~ and \BBs~ mixings, and we have presented a detailed account of the individual impact that they may have in the analysis of CP violation and flavour physics when one allows for NP contributions to those mixings. Figure \ref{figure11} shows some important results of a complete analysis that incorporates the new constrains provided by their measurements. The complete set of observables is:
\begin{eqnarray*}
|\V{ud}|\quad\quad |\V{us}|\quad\quad |\V{ub}|\quad\quad |\V{cd}|\quad\quad |\V{cs}|\quad\quad |\V{cb}|\\
\AJPsi\quad\quad \gamma \quad\quad \bar\alpha \quad \quad 2\bar\beta+\gamma\quad\quad\cos 2\bar\beta \quad\quad \DMBd \\
\DMBs\quad\quad\Asld\quad\quad \DGBd \quad\quad \Ads \quad \quad \DGBsCP.\label{completeset}
\end{eqnarray*}

The numerical values are shown in table \ref{table:inputs}.
At this stage, the following comments are in order.
\begin{itemize}
\item The SM-like solution $\gamma\sim 65^\circ$ almost emerges as the only relevant one with 97.3\% probability. There is still some room for the NP solution with $\gamma\sim -115^\circ$, although it retains 2.6\% probability. $\Asld$ and $\Ads$ are the observables sensitive to $\gamma$ vs. $\gamma+\pi$. Their combined action (see figures \ref{figure02b} and \ref{figure08a}) is the cause of the reduction of probability for the NP solution. For an explicit example of a NP model -- in this case the Littlest Higgs Model with T-Parity -- allowing $\gamma\sim -115^\circ$ and showing some model independent features like the importance of $\Asld$ addressed here, see reference \cite{Blanke:2006sb}.
\item The available observables involving the \BBs~ system are not useful to distinguish $\gamma$ from $\gamma+\pi$ solutions. This is ultimately connected to the hierarchical nature of the CKM matrix elements and the flatness of the unitarity triangle $sb$, the relevant one for \BBs, in constrast to the \BBd~ case. In this sense, model independent NP studies of the present kind in both sectors are \emph{decoupled}.
\item Recent determinations of $\DMBs$, being within SM expectations, force $r_s\sim 1$; this fact makes an important difference with respect to $r_d$, which can have values significantly different from 1 in NP scenarios. This is an interesting property since in some specific NP models, those two quantities may be related. The combined knowledge of $\DMBd$, $\DMBs$ can provide an important tool to discriminate models (see figure \ref{figure11b}).
\item $\Asls$, paralleling the r\^{o}le of $\Asld$ in $B_d$'s, is sensitive to the phase of the mixing \BBs, $2\bar\chi$, and can thus provide some information on $2\phi_s$; it is not, however, accessible to experiment in hadronic machines, and its usefulness is transferred to $\Ads$. $\Ads$, being a mixture of $\Asld$ and $\Asls$, is sensitive to both $\phi_d$ and $\phi_s$. The present uncertainty in its measurement limits, however, its constraining ability to favouring $2\phi_d\sim 0$ and $2\phi_s>0$.
\item The actual measurement of the width $\DGBd$ is not useful in providing any effective constraint.
\item In opposition to $\DGBd$, $\DGBs$ or more precisely, what is really measured, $\DGBsCP$, does provide some information on $\phi_s$, and it favours values producing large $\cos^2(2\bar\chi)\simeq\cos^2(2\phi_s)$.
\item The absence of measurements directly testing the phase in \BBs~ mixing, as for example $A_{J/\Psi \Phi}$, allows the NP parameter $\phi_s$ to be relatively free, up to the reduced constraining power of $\Ads$ and $\DGBsCP$. This can be seen by comparing figures \ref{figure06a} and \ref{figure11d}. Figure \ref{figure12} shows the probability distribution of $2\bar\chi$ after using the complete set of available constraints, including $\Ads$ and $\DGBsCP$ which are the only ones that give some information on $\phi_s$. The result can be well understood in terms of what could be expected from figures \ref{figure08b} and \ref{figure09b}: $\Ads$ favours positive, large values of $\sin 2\bar\chi\simeq \sin 2\phi_s$ while $\DGBsCP$ favours large values of $\cos^2(2\bar\chi)\simeq\cos^2(2\phi_s)$; figure \ref{figure12} reflects the interplay among both constraints. From this distribution it is hard to give a prediction for $A_{J/\Psi \Phi}$ except that with the NP considered here large negative values of $A_{J/\Psi \Phi}$ are more disfavoured.

\begin{figure}[h]
\begin{center}
\epsfig{file=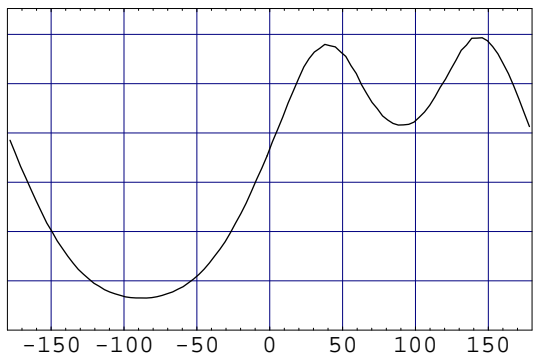,width=0.55\textwidth}
\caption{Probability distribution, $2\bar\chi$.}\label{figure12}
\end{center}
\end{figure}
\end{itemize}

\begin{figure}[htb]
\begin{center}
\subfigure[$2\phi_d$\label{figure13a}]{\epsfig{file=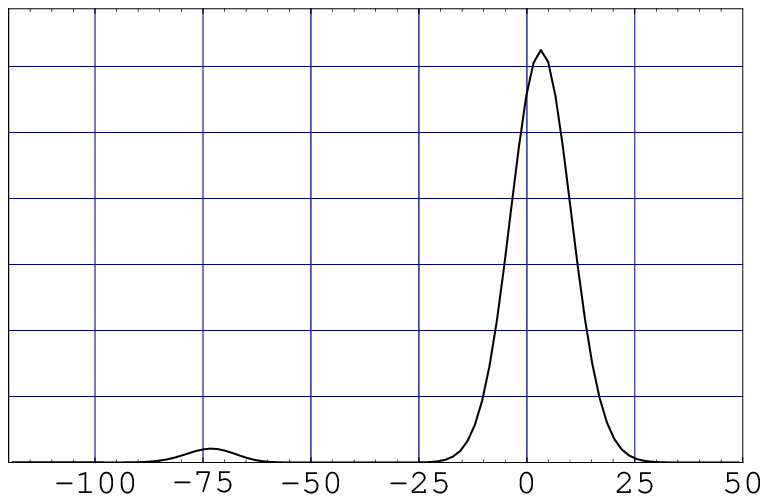,width=0.45\textwidth}}\quad \subfigure[$r_d^2$\label{figure13b}]{\epsfig{file=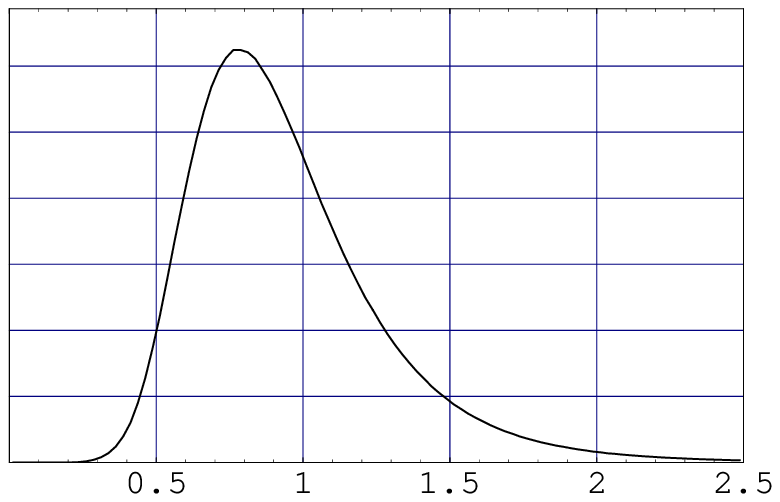,width=0.45\textwidth}}\\
\subfigure[$r_s^2$\label{figure13c}]{\epsfig{file=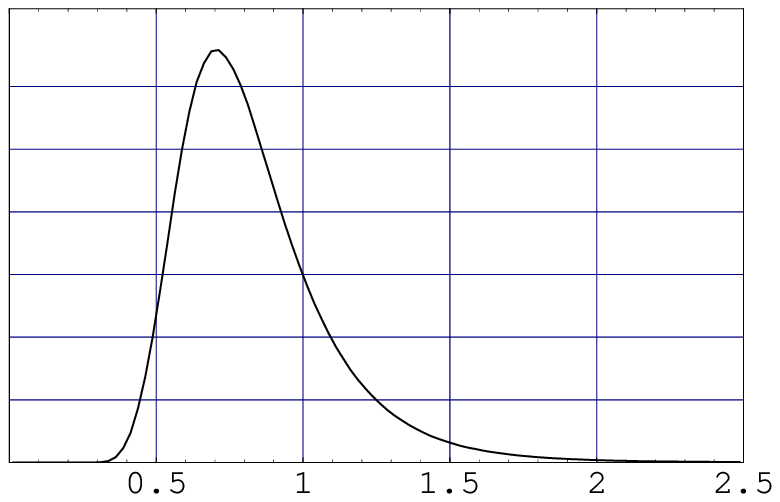,width=0.45\textwidth}}\quad \subfigure[$r_d^2/r_s^2$\label{figure13d}]{\epsfig{file=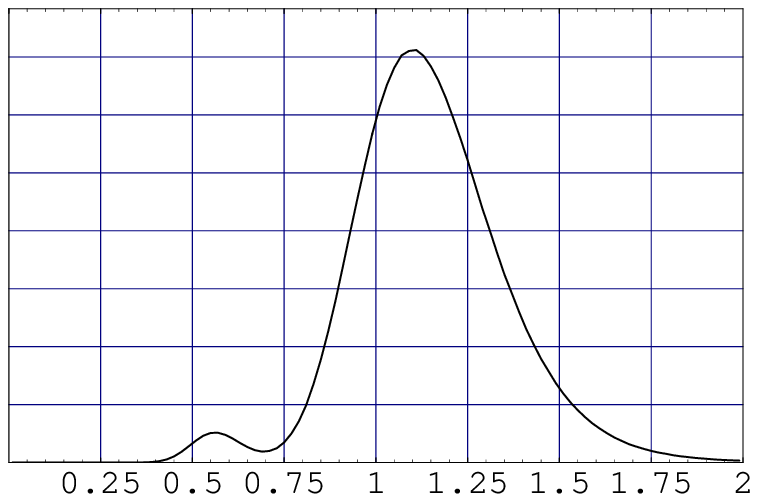,width=0.45\textwidth}}\\
\subfigure[$h_d$ vs. $\sigma_d$\label{figure13e}]{\epsfig{file=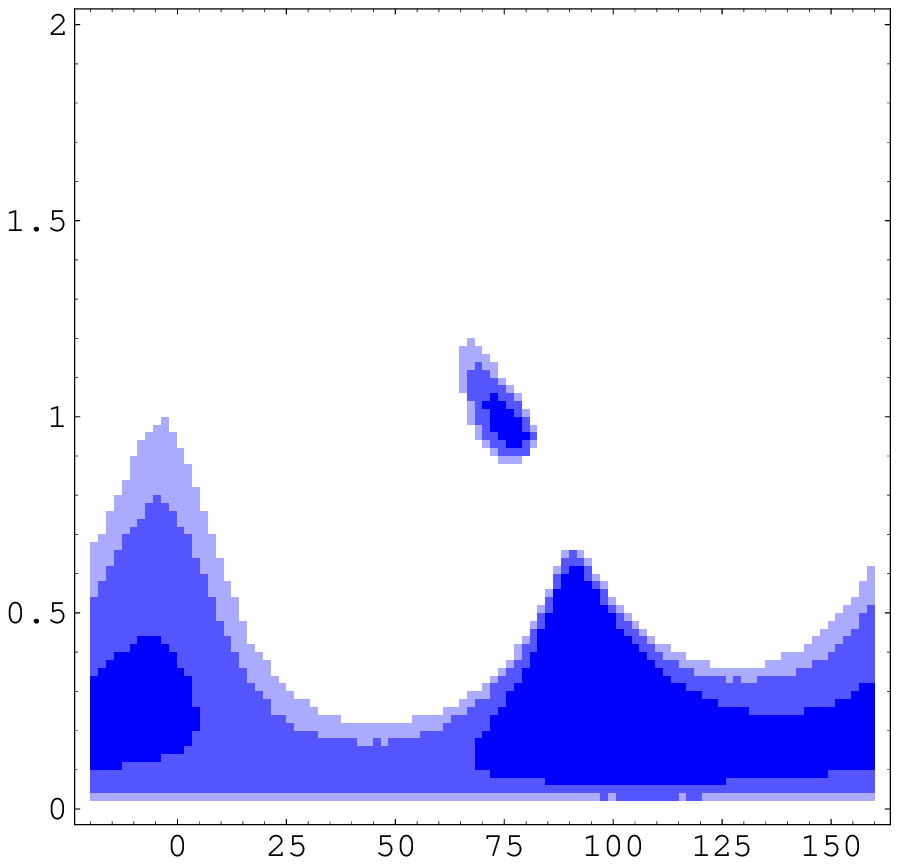,width=0.45\textwidth}}\quad \subfigure[$h_s$ vs. $\sigma_s$\label{figure13f}]{\epsfig{file=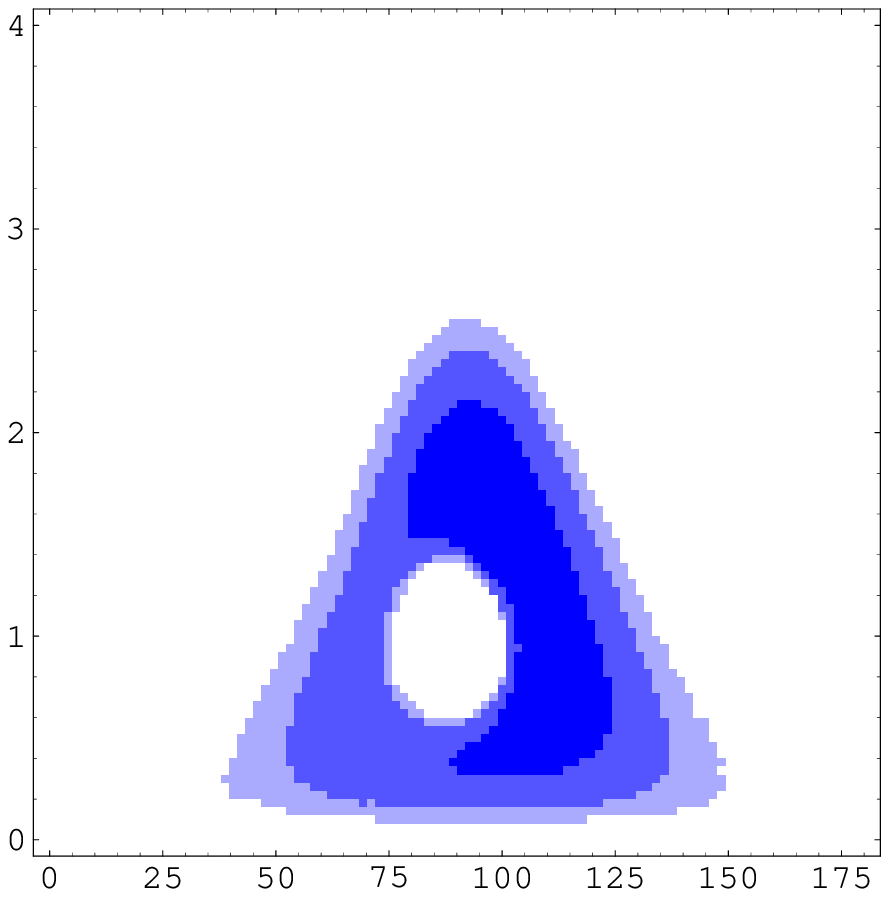,width=0.44\textwidth}}
\caption{Probability distributions.}\label{figure13}
\end{center}
\end{figure}

Figure \ref{figure13} includes additional results of the analysis. The probability distribution of $2\phi_d$ in figure \ref{figure13a} shows the SM-like solution $2\phi_d=(3\pm 7)^\circ$ and the still present NP solution with $2\phi_d=(-75\pm 7)^\circ$. Concerning $r_d^2$, they correspond, respectively, to $r_d^2=0.51\pm 0.16$ and $r_d^2=0.97\pm 0.32$, and merge into the distribution in figure \ref{figure13b} with $r_d^2=0.94\pm 0.33$. The probability distribution of $r_s^2$ in figure \ref{figure13c} yields $r_s^2=0.84\pm 0.26$. In figure \ref{figure13d} the probability distribution of $r_d^2/r_s^2$ exhibits the presence of both solutions, $r_d^2/r_s^2=0.56\pm 0.08$ and $r_d^2/r_s^2=1.17\pm 0.19$.  Finally, figures \ref{figure13e} and \ref{figure13f} contain the same information as figures \ref{figure11c} and \ref{figure11d} in a different parametrisation: $r_q^2e^{-i2\phi_q}=1+h_qe^{i2\sigma_q}$, $q=d,s$. 

The general trend of the previous results is in rather good agreement with references \cite{Ligeti:2006pm,Ball:2006xx,Grossman:2006ce,Bona:2006sa}, which typically use the same kind of observables to explore the allowed NP parameter space. For comparison we will comment on some minor differences with similar results in reference \cite{Bona:2006sa}. Our $r_d^2$ distribution has a smaller width due to the difference in the $\xi$ input. The small valley around $90^\circ$ in the distribution of figure \ref{figure12} is much deeper than the one in the corresponding figure of \cite{Bona:2006sa}. The recent shift in the measurement of $\Ads$ by the D0 collaboration \cite{D0ASL} is at the origin of this difference as can be easily understood from our figure \ref{figure08a}. One cannot say that the D0 measurement suggests NP but the interplay between $\Ads$ and $\DGBsCP$ is becoming quite interesting. Finally, the NP solution with $\gamma$ in the third quadrant has a 1\% probability in reference \cite{Bona:2006sa} in contrast with our 2.6\%. Taking into account the differences in hadronic inputs and other different inputs as $\bar\alpha$ -- see appendix \ref{sec:inputs} -- this difference should be regarded as non-conflicting.

\section{Summary and Conclusions\label{sec:Conclusions}}
We analyse in a systematic way the constraints on the SM and on New Physics which are implied by the presently available information on $B_d$ and $B_s$ systems. Assuming that tree level meson decays are dominated by the SM amplitudes, but allowing for New Physics contributions to \BBd~ and \BBs~ mixings, we analyse in detail the various solutions which are still allowed by data. Our analysis includes a detailed study of the impact which each individual measurement has in shaping up the allowed regions for New Physics. This is specially relevant to gauge the importance that improved experimental results will have on the prospects to either keep New Physics contributions adjacent to the SM or to allow clear differences. The main NP solution with $\gamma\sim -115^\circ$ and $2\phi_d\sim -75^\circ$ still retains $2.6\%$ probability. As stressed, this relative suppression is mainly due to the central r\^ole played by $\text{Im}[\Gamma_{12}^d/ M_{12}^d]$ in the semileptonic asymmetries $\Asld$ and $\Ads$. The actual small number of available observables and their particular dependence on the phase of the mixing, $2\bar\chi$, leaves ample room for New Physics in the \BBs~ system.

\section*{Acknowledgments}
This research has been supported by European FEDER, Spanish MEC under grant FPA 2005-01678, \emph{Generalitat Valenciana} under GVACOMP 2006-214, by \emph{Funda\c{c}\~{a}o para a Ci\^{e}ncia e a Tecnologia} (FCT, Portugal) through the projects POCTI/FNU/44409/2002, PDCT/FP/FNU/50250/2003, POCI/FP/\\63415/2005, POCTI/FP/FNU/50167/2003, CFTP-FCT UNIT 777, and by \emph{Ac\c{c}\~{a}o Integrada Luso-Espanhola} E73/04. The authors acknowledge A. Buras, C.S. Kim and D. Zieminska for useful comments. G.C.B. is grateful to A. Buras for kind hospitality at TUM. M.N. acknowledges financial support from FCT.

\appendix
\section{Numerical framework \& Inputs \label{sec:inputs}}
The statistical framework we have followed in this paper is bayesian analysis, which yields probability distributions obtained from the interplay between the constraints imposed by experimental measurements and prior knowledge (or ignorance) on the parameters. At the more technical level, we make use of Markov Chain driven MonteCarlo simulations to calculate the considered probability distributions. The typical size of the corresponding random walks, i.e. the number of points in each distribution, is $2\times 10^8$ steps/points.

Table \ref{table:inputs} summarises the most relevant quantities used in the analyses of the previous sections. The numerical values come from a variety of sources that include the Particle Data Group \cite{PDG}, the Heavy Flavour Averaging Group \cite{unknown:2006bi} and results from the Babar \cite{Aubert:2004zt,Sciolla:2005kz,Aubert:2004kv,Aubert:2006ia,Aubert:2003wr,Aubert:2004zr,Aubert:2005nj,Aubert:2006af,Aubert:2004cp,Aubert:2005yf,Aubert:2006tw}, Belle \cite{Abe:2005bt,Abe:2004gu,Poluektov:2006ia,Zhang:2003up,Somov:2006sg,Itoh:2005ks,Abe:2004mu} -- concerning B factories --, D0 \cite{Abazov:2006dm,D0ASL,D0DGBs} and CDF \cite{Abulencia:2006mq,CDFDGBs} -- concerning B physics at Tevatron --  collaborations. The measured value of $\bar\alpha$ deserves some comment. There is no complete agreement on the value and the shape of this constraint among different authors (see \cite{Charles:2006vd}) and thus, following the discussion in \cite{Botella:2005fc} concerning it, in the present work we use a pair of gaussian distributions centered at $-80^\circ$ and $100^\circ$ respectively, with standard deviations $11^\circ$. For simplicity, all uncertainties are modeled as gaussians.

\begin{table}[h]
\begin{center}
\begin{tabular}{|c|c||c|c|}
\hline
$|\V{ud}|$ & $0.9738\pm 0.0005$  & $|\V{us}|$ & $0.2200\pm 0.0026$\\ \hline
$|\V{cd}|$ & $0.224\pm 0.012$ & $|\V{cs}|$ & $0.976\pm 0.013$\\ \hline
$|\V{ub}|$ & $(40\pm 4)10^{-4}$ & $|\V{cb}|$ & $(41.3\pm 1.5)10^{-3}$\\ \hline
$\AJPsi$ & $0.674\pm 0.026$ & $\gamma$ & $(-115,65\pm 18)^\circ$\\ \hline
$\bar\alpha$ & $(-80,100\pm 11)^\circ$ & $2\bar\beta+\gamma$ & $(-90,90\pm 46)^\circ$\\ \hline
$\cos 2\bar\beta$ & $1.9\pm 1.3$ &  & \\ \hline
$\DMBd$ & $(0.507\pm 0.005)$ ps$^{-1}$ & $\DMBs$ & $(17.4\pm 0.4)$ ps$^{-1}$\\ \hline
$\Asld$ & $-0.003\pm 0.0078$ & $\Ads$ & $-0.0028\pm 0.0016$\\ \hline
$\DGBsCP$ & $(0.15\pm 0.10)$ ps$^{-1}$ & $\DGBd/\Gamma_d$ & $0.009\pm 0.037$\\ \hline
$\xi$ & $1.24\pm 0.04$  & $\sqrt{B_{B_s}}f_{B_s}$ & $(0.276\pm 0.038)$ GeV\\ \hline

\end{tabular}
\caption{Inputs}\label{table:inputs}
\end{center}
\end{table}

\newpage


\end{document}